%% file: ISIT2019_long.tex
\documentclass[10pt, draftclsnofoot, onecolumn,romanappendices]{IEEEtran}

	\usepackage[utf8]{inputenc} 
	\usepackage[T1]{fontenc}
	\usepackage{url}
	\usepackage{ifthen}
	\usepackage{cite}
	\usepackage[cmex10]{amsmath} % Use the [cmex10] option to ensure complicance
	\usepackage{pgfplots}
	\usepackage{bm}
	\usepackage{amsbsy}
	\usepackage{epsfig} 
	\usepackage{latexsym} 
	\usepackage{amssymb}
	\usepackage{wasysym}
	\usepackage{placeins}
	\usepackage{color}
	\usepackage{rotating}
	\usepackage{dsfont}
	\usepackage{caption}
	\usepackage{subcaption}
	\usepackage{algorithm}
	\usepackage{graphicx}
	\usepackage[shortlabels]{enumitem}
	\usepackage{epstopdf}
	\usepackage{cancel}
	\usepackage{algpseudocode}
	\usepackage{units}
	\usepackage{multirow}
	\usepackage{cases}
	\usepackage{tabularx}

\usetikzlibrary{shapes,arrows,snakes,positioning}
\usetikzlibrary{decorations.markings}
\usetikzlibrary{calc}
\usepackage[ddmmyyyy]{datetime}

\usepackage{csquotes}
\usepackage{tikz}

\usepackage[percent]{overpic}

	\usepackage{hyperref}

	\input{./Definitions}

\newtheorem{theorem}{Theorem}
\newtheorem{definition}{Definition}

\newtheorem{lemma}{Lemma}
\newtheorem{corollary}{Corollary}

\begin{document}
	\title{Achieving Vanishing Rate Loss in Decentralized Network MIMO}
	\author{%
		\IEEEauthorblockN{Antonio Bazco-Nogueras\IEEEauthorrefmark{1}\IEEEauthorrefmark{2},
											Lorenzo Miretti\IEEEauthorrefmark{2}, 
											Paul de Kerret\IEEEauthorrefmark{2}, 
											David Gesbert\IEEEauthorrefmark{2}, 
											Nicolas Gresset\IEEEauthorrefmark{1}}
		\IEEEauthorblockA{\IEEEauthorrefmark{1}%
											Mitsubishi Electric R\&D Centre Europe (MERCE),
											35700 Rennes, France,
											\{a.bazconogueras,n.gresset\}@fr.merce.mee.com}
		\IEEEauthorblockA{\IEEEauthorrefmark{2}%
											Communication Systems Department, EURECOM,
											06410 Biot, France,
											\{bazco,miretti,dekerret,gesbert\}@eurecom.fr}
	}

 %\date{\today}
	\maketitle

	\begin{abstract} 
			%In this paper\footnote{D. Gesbert, P. de Kerret, and L. Miretti are supported by the European Research Council under the European Union's Horizon 2020 research and innovation program (Agreement no. 670896).}, we analyze the Network MIMO channel with 2 Transmitters (TXs) jointly serving 2 users, where each TX has a different multi-user Channel State Information (CSI), each one with different accuracy. 
			%It is known that this decentralized setting attains the DoF of the centralized counterpart in which both TXs share the best CSIT. However, the rate gap between both settings has remained unbounded. By means of the affine approximation of the capacity at high-SNR, we unexpectedly show that it is possible to recover asymptotically the sum rate achieved by Zero-Forcing (ZF) precoding in the centralized setting with perfect CSI sharing between TXs. This is achieved through a novel precoding scheme which manages to enforce at the same time a high precision ZF precoding and an accurate instantaneous decentralized power control.
			
			%%%%%%%%%%%%%% David corrections %%%%%%%%%%%%%%
			
			In this paper\footnote{L. Miretti, P. de Kerret, and D. Gesbert are supported by the European Research Council under the European Union's Horizon 2020 research and innovation program (Agreement no. 670896 PERFUME).},
			we analyze a Network MIMO channel with 2 Transmitters (TXs) jointly serving 2 users, where each TX has a different multi-user Channel State Information (CSI), potentially with %each one with 
			a different accuracy. 
			Recently it was shown the surprising result that this decentralized setting can attain the same Degrees-of-Freedom (DoF) as its genie-aided centralized counterpart in which both TXs share the best-quality CSI. 
			However, the DoF derivation alone does not characterize the actual rate and the question was left open as to how big the rate gap between the centralized and the decentralized settings was going to be.
			%However, the rate gap between both settings has remained unbounded. 
			%By means of the affine approximation of the capacity at high-SNR, 
			In this paper, we considerably strengthen the previous intriguing DoF result by showing that it is possible %we unexpectedly show that it is possible 
			to achieve asymptotically the same sum rate as that attained by Zero-Forcing (ZF) precoding in a centralized setting endowed with the best-quality CSI. %with perfect CSI sharing among TXs. 
			This result involves a novel precoding scheme which is tailored to the decentralized case. The key intuition behind this scheme lies in the striking of an asymptotically optimal compromise between i) realizing high enough precision ZF precoding while ii) maintaining consistent-enough precoding decisions across the non-communicating cooperating TXs. 

			%This is achieved through a novel precoding scheme which manages to enforce at the same time a high precision ZF precoding and an asymptotically optimal beamforming gain through highly consistent decisions. 
			
			%%%%%%%%%%%%%% %%%%%%%%%%%%%% %%%%%%%%%%%%%%%%%			
	\end{abstract} 
	
	%%%%%%%%%%%%%%%%%%%%%%%%%%%%%%%%%%%%%%%%%%%%%%%%%%%%%
	%%%%%%%%%%%%%%%%%%%%%%%%%%%%%%%%%%%%%%%%%%%%%%%%%%%%
	%%%%%%%%%%%%%%%%%%%%%%%%%%%%%%%%%%%%%%%%%%%%%%%%%%%%%
		\section{Introduction}\label{se:intro}
	%%%%%%%%%%%%%%%%%%%%%%%%%%%%%%%%%%%%%%%%%%%%%%%%%%%%%
	%%%%%%%%%%%%%%%%%%%%%%%%%%%%%%%%%%%%%%%%%%%%%%%%%%%%%
	%%%%%%%%%%%%%%%%%%%%%%%%%%%%%%%%%%%%%%%%%%%%%%%%%%%%%

			Joint transmission in wireless networks is known to bring multiplicative improvements in network rates only under the assumption of perfect CSI\cite{Davoodi2016_TIT_DoF}. 
			The study of how imperfect or quantized CSI at the TXs (CSIT) affects the performance %of this technique 
			has focused on the assumption that the imperfect information is \emph{perfectly shared} between the non-colocated transmitting antennas\cite{Davoodi2016_TIT_DoF,Jindal2006}. %, either assuming a single TX or a perfect backhaul between several TXs.  
			However, this assumption may not be adapted to many applications within the upcoming wireless networks use cases,  such as Ultra-Reliable Low-Latency Communication (URLLC) or heterogeneous backhaul deployments. 
			As a result, there is a clear interest in looking at the %Consequently, the interest about the 
			scenario in which each TX may have a different information about the channel, denoted as \emph{Distributed CSIT} setting\cite{Gesbert2018_Chapterss}. %,  has raised in the last years. 
			
			We focus in this paper on a particular sub-case of the Distributed CSIT setting, so-called Distributed Network MIMO, wherein
			%on the Distributed Network MIMO setting, in which 
			the TXs have access to all the information symbols of the users (RXs), yet do not share the same CSIT\cite{dekerret2012_TIT}. 
			This model arises in presence of caching \cite{MaddahAli2014} and Cloud-RAN with high mobility\cite{Peng2015}, in which latency constraints impede efficient CSIT sharing within the channel coherence time.			
			The DoF of this scenario has been studied in previous works. 
			Specifically, it was shown that conventional ZF performs very poorly and several  schemes were proposed to improve the robustness of the transmission with respect to CSIT inconsistencies\cite{dekerret2012_TIT,Bazco2018_WCL,Bazco2018_TIT}. 
			One of the main successes has been obtained for the 2-user setting where the DoF was shown to be equal to the DoF of the centralized setting\cite{Bazco2018_WCL} through an asymmetric precoding scheme where some TX deliberately throws away %the TXs may not exploit
			 instantaneous CSIT.
			
			Yet, these works suffer from the limitations of the metric used, as DoF only provides the asymptotic rate \emph{slope} with respect to the SNR.
			Since it does not provide any information about the beamforming gain or the efficient power use at the TXs, schemes resulting in the same DoF may need a considerably different power to achieve the same rate\cite{Lozano2005}. 
			%achievable rate for a given SNR and, as shown in\cite{Lozano2005}, schemes resulting in the same DoF may need a considerably different power to achieve the same rate. 
			Hence, the natural next step towards capacity characterization is 
			to study the \emph{rate offset}, which is the constant term in the linear approximation of the sum rate at high SNR. %, i.e., %to analyze the capacity as an affine function of the SNR. 
			%This capacity approximation allows for an asymptotic analysis of the achievable rate and  writes as
			Based on this linear approximation, the rate expression can be written as \cite{Lozano2005}
				\eqm{
						R(P) = \DoF\log_2(\SNR) - \Lc_\infty + o(1) \label{eq:def_affine}
				}
			where $\Lc_\infty$ represents the {rate offset} (vertical offset) and $\limpf o(1) = 0$. %The \emph{rate offset} (vertical offset)  is directly obtained  as $\Lc_\infty/\DoF$. 
			%
			%The second important limitation comes from the average power constraint considered in \cite{dekerret2012_TIT,Bazco2018_WCL,Bazco2018_TIT}. Such a constraint is not practical and is not in line with conventional assumptions. Considering an instantaneous power constraint, all the schemes presented in \cite{dekerret2012_TIT,Bazco2018_WCL,Bazco2018_TIT} do not achieve the optimal DoF. The only scheme achieving the optimal DoF is obtained from \cite{dekerret2012_TIT} where the transmit power scales in ${P}/{\log(P)}$ for a maximum instantaneous power of $P$. This leads to a very inefficient power control, and hence to a very poor rate offset.
						%
						%						
			%On the basis of this affine analysis of the capacity, we analyze the high-SNR regime of the 2x2 Distributed Network MIMO setting. 
			Our main contributions read as follows:
				\begin{itemize}
						\item We provide a novel precoding scheme that achieves accurate ZF of the interference and, at the same time, a high beamforming gain through consistent transmission at the TXs.
						\item Through a new lower bound, we show that the proposed scheme achieves a vanishing rate loss at high SNR when compared to the centralized configuration with perfect CSIT sharing.
				\end{itemize}
				
			%%%%%%%%%%%%%%%%%%%%%%%%%%%%%%%%%%%%%%%%%%%%%%%%%%%%%
			\paragraph*{Notations} 
					We use the Landau notation, i.e.,  $f(x) = o(x)$ implies than $\lim_{x\rightarrow\infty}\!\frac{\!f(x)\!}{x}=0$, and $f(x) = O(x)$ implies than $\lim_{x\rightarrow\infty}\!\frac{\!f(x)\!}{x}=M$, with $0<|M|<\infty$. $\Rb^+$ stands for $\{x\in\Rb:x>0\}$, and $\Eb_{|A}$ denotes  the conditional expectation given an event $A$. 
					%			
					%We use $\doteq$ to denote exponential equality, i.e., we write $f(P) \doteq P^x$ to denote $\lim_{P\rightarrow \infty} \frac{\log f(P)}{\log P}= x$. 
					%The exponential inequalities $\dotleq$ and $\dotgeq$ are defined in the same way. Similarly, w
					%We use the Landau notation, i.e., $f(x) = O(x)$ implies than $0 < \lim_{x\rightarrow\infty}\frac{f(x)}{x} < \infty$, and $f(x) = o(x)$ implies than $\lim_{x\rightarrow\infty}\frac{f(x)}{x}=0$. 
					%We denote the Hadamard (element-wise) product of two matrices $\bA$ and $\bB$ as $\bA\odot \bB$. 
					%The determinant of a square matrix $\bA$ is denoted by $\det(\bA)$. 
					%The Hermitian transpose of a matrix $\bA$ is denoted as $\bA^\He$. 
					%The unitary circularly-symmetric complex normal distribution is denoted as  $\CN(0,1)$.
					$\norm{\bv}$ stands for the Euclidean norm of the vector $\bv$, and $\Pr(A)$ denotes the probability of the event $A$.  
					%Given a vector $\bv$, $\tilde{\bv}$ is defined as $\tilde{\bv} = \frac{{\bv}}{\norm{\bv}}$.  % and $\angle(\bv,\bw)$ denotes the angle between the vectors $\bv$ and $\bw$, such that $\abs{\cos(\angle(\bv,\bw))}=\frac{\abs{\bv^\He\bw}}{\norm{\bv}\norm{\bw}}$.
					%For any $z\in\Cb$, $\conj(z)$ denotes the complex conjugate of $z$.
					%$\Rb^+$ denotes the set of strictly positive real numbers.
					%The first $m$ positive integers are denoted by $\Nb_m=\{1,\dots, m\}$
			%%%%%%%%%%%%%%%%%%%%%%%%%%%%%%%%%%%%%%%%%%%%%%%%%%%%%

	%%%%%%%%%%%%%%%%%%%%%%%%%%%%%%%%%%%%%%%%%%%%%%%%%%%%%%%%%%%%%%%%%%%%%%%%%%%%%%%%%
	%%%%%%%%%%%%%%%%%%%%%%%%%%%%%%%%%%%%%%%%%%%%%%%%%%%%%%%%%%%%%%%%%%%%%%%%%%%%%%%%%
	%%%%%%%%%%%%%%%%%%%%%%%%%%%%%%%%%%%%%%%%%%%%%%%%%%%%%%%%%%%%%%%%%%%%%%%%%%%%%%%%%
	%%%%%%%%%%%%%%%%%%%%%%%%%%%%%%%%%%%%%%%%%%%%%%%%%%%%%%%%%%%%%%%%%%%%%%%%%%%%%%%%%
		\section{Problem Formulation}\label{se:SM}
	%%%%%%%%%%%%%%%%%%%%%%%%%%%%%%%%%%%%%%%%%%%%%%%%%%%%%%%%%%%%%%%%%%%%%%%%%%%%%%%%%
	%%%%%%%%%%%%%%%%%%%%%%%%%%%%%%%%%%%%%%%%%%%%%%%%%%%%%%%%%%%%%%%%%%%%%%%%%%%%%%%%%
	%%%%%%%%%%%%%%%%%%%%%%%%%%%%%%%%%%%%%%%%%%%%%%%%%%%%%%%%%%%%%%%%%%%%%%%%%%%%%%%%%
	%%%%%%%%%%%%%%%%%%%%%%%%%%%%%%%%%%%%%%%%%%%%%%%%%%%%%%%%%%%%%%%%%%%%%%%%%%%%%%%%%

		%%%%%%%%%%%%%%%%%%%%%%%%%%%%%%%%%%%%%%%%%%%%%%
		%%%%%%%%%%%%%%%%%%%%%%%%%%%%%%%%%%%%%%%%%%%%%%
			\subsection{Transmission Model}
		%%%%%%%%%%%%%%%%%%%%%%%%%%%%%%%%%%%%%%%%%%%%%%
		%%%%%%%%%%%%%%%%%%%%%%%%%%%%%%%%%%%%%%%%%%%%%%		
			We consider a setting with $2$~single-antenna TXs jointly serving $2$~single-antenna RXs over a Network MIMO setting --also known as Distributed Broadcast Channel (BC)--. 
			%The extension from the single-antenna setting to the setting with $N$~antennas per node follows easily but it is not presented here for the sake of clarity. 
			The extension to a setting with multiple-antenna nodes in which every TX and RX has the same number of antennas $N$ follows naturally. Yet, it comes at the cost of heavier and less intuitive notations such that we focus here on the single antenna case. The extension to multiple-antenna TXs but single-antenna RXs is more challenging and is relegated to the journal version of this work. 
			The signal received at RX~$i$ is % writes as
				\begin{equation}
						y_i=\hv_i^{\He}\xv+z_i,	 	\label{eq:SM_1}
				\end{equation}
			where $\hv_i^{\He}\in \Cb^{1\times 2}$ is the channel coefficients vector towards RX~$i$, ${\xv\in\mathbb{C}^{2\times 1}}$ is the transmitted multi-user signal, and $z_i\in \mathbb{C}$ is the Additive White Gaussian Noise (AWGN) at RX~$i$, being independent of the channel and the transmitted signal, and drawn from a circularly symmetric complex Gaussian distribution~$\CN(0,1)$. We further define the channel matrix~$\bH\in \Cb^{2\times 2}$ as
				\begin{equation}
						\bH\triangleq \begin{bmatrix}
									\hv_1^{\He}\\
									\hv_2^{\He}
							\end{bmatrix},			\label{eq:SM_2}
				\end{equation}
			with its $(i,k)$-th element representing the channel coefficient from TX~$k$ to RX~$i$ and being denoted as $\h_{ik}$. 
			The channel coefficients are assumed to be %independently and identically distributed 
			i.i.d. as $\CN(0,1)$ 
			such that all the channel sub-matrices are full rank with probability one.
						
			The transmitted multi-user signal~$\xv \in \Cb^{2\times 1}$, is obtained from the precoding of the symbol vector $\bs  \triangleq [s_1\ s_2]^{\Transpose}$. The symbols $s_i$ are i.i.d. as $\CN(0,1)$ and $s_i$ denotes the symbol intended by RX~$i$ such that %\vspace{-.5ex} 						
			%The transmitted multi-user signal~$\xv \in \Cb^{2\times 1}$, is obtained from the precoding of the symbol vector $\bs  \triangleq [s_1\ s_2]^{\Transpose}$, where the symbols $s_i$ are i.i.d. as $\CN(0,1)$ and $s_i$ denotes the symbol intended by RX~$i$ such that 
				\begin{equation}
						\xv  \triangleq	{\Pb} \begin{bmatrix}\tv_{1} & \tv_{2}\end{bmatrix} 		\begin{bmatrix} s_1 \\ s_2 \end{bmatrix},   				\label{eq:SM_3}
				\end{equation}
			where $\Pb\triangleq \sqrt{P}$ and $P$ is the maximum transmit power per TX. 
			The vector $\tv_{i}\in\Cb^{2\times 1}$ denotes the normalized precoding  vector towards RX~$i$. 
			For further reference, we also introduce the multi-user precoder~$\bT\in \mathbb{C}^{2\times 2}$ as $\bT\triangleq \begin{bmatrix}\tv_{1} & \tv_{2}\end{bmatrix}$, and the precoder of TX~$j$ as  $\tv_{\TX j}\triangleq \begin{bmatrix}\{\tv_{1}\}_{j} & \{\tv_{1}\}_{2}\end{bmatrix}{}^{\!\Trans}$. 			
			We assume  a per-TX instantaneous power constraint  \emph{for the precoder}, i.e., 
			$\norm{\tv_{\TX j}}\leq 1$, $\forall j\in\{1,2\}$, such that $\ExpB{\norm{\xv}^2}\leq P$. 
			%such that $\norm{\tv_{\TX j}}\leq 1$, $\forall j\in\{1,2\}$. 
		%%%%%%%%%%%%%%%%%%%%%%%%%%%%%%%%%%%%%%%%%%%%%%
		%%%%%%%%%%%%%%%%%%%%%%%%%%%%%%%%%%%%%%%%%%%%%%
			\subsection{Grassmanian Random Vector Quantization}
		%%%%%%%%%%%%%%%%%%%%%%%%%%%%%%%%%%%%%%%%%%%%%%
		%%%%%%%%%%%%%%%%%%%%%%%%%%%%%%%%%%%%%%%%%%%%%%
			We consider in this work that the RXs have perfectly estimated the channel coefficients to focus on the challenges of CSI feedback  and limited CSI sharing among TXs. 
			As we analyze the high SNR performance, we follow the same approach of the reference work from Jindal \cite{Jindal2006} and study the performance of Grassmanian Random Vector Quantization (RVQ). 
			For the sake of completeness, we will recall in the following some properties that will be needed in the proof of our main results. For more details about RVQ, see~\cite{Au-Yeung2007,Jindal2006}.
			
			In RVQ, a unit-norm channel vector~$\tilde{\hv}\in \mathbb{C}^M$ is quantized using $B$~bits to a codebook~$\mathcal{C}$ containing~$2^B$ unit-norm vectors isotropically distributed on the $M$-dimensional unit sphere.
			We consider a Grassmanian quantization scheme such that the quantized estimate denoted by~${\bhh}\in \mathbb{C}^M$ is obtained to minimize the angle with the true channel, i.e.,
				\begin{equation}
					\begin{aligned}
							{\bhh} 
									&=\argmax_{\bw\in \Cc} |\tilde{\hv}^{\He}\bw|^2\\
									&=\argmin_{\bw\in \Cc} \sin^2(\measuredangle(\tilde{\hv},\bw)),
					\end{aligned}			\label{eq:SM_4}
				\end{equation}  
			where we have introduced the angle for unit-norm vectors in~$\Cb^{M}$ from $\measuredangle(\bx,\by)\triangleq \arccos |\bx^{\He}\by|$. We define the quantization error as 
				\begin{equation}
						Z\triangleq \sin^2(\tilde{\hv},\bhh).
				\end{equation}
			Since the elements of the codebook~$\mathcal{C}$ are independent of~$\tilde{\hv}$ and isotropically distributed, the quantization error~$Z$ is obtained as the minimum of $2^B$ independent beta $(M-1,1)$ random variables. Upon defining $z=\sqrt{Z}$, and $\zop\triangleq \sqrt{1-Z}$, we can write the true channel as a function of its quantized version as
				\begin{equation} 
						\thv= \zop\bhh+ z\bm{\delta},			\label{eq:SM_err2}
				\end{equation}
			where $\bm{\delta}$ is a unit-norm vector isotropically distributed in the null space of $\hat{\bh}$, and $\bm{\delta}$ and $Z$ are mutually independent. In our setting, since the vectors have $M=2$ elements, the quantization error $Z$ is distributed as the minimum of $2^B$~standard uniform random variables \cite{Jindal2006}.

		%%%%%%%%%%%%%%%%%%%%%%%%%%%%%%%%%%%%%%%%%%%%%%
		%%%%%%%%%%%%%%%%%%%%%%%%%%%%%%%%%%%%%%%%%%%%%%
			\subsection{Distributed CSIT Model}
		%%%%%%%%%%%%%%%%%%%%%%%%%%%%%%%%%%%%%%%%%%%%%%
		%%%%%%%%%%%%%%%%%%%%%%%%%%%%%%%%%%%%%%%%%%%%%%
			%We consider in this work that the RXs have perfect channel knowledge to focus on the challenges of CSI feedback and limited CSI sharing among TXs. 
			As previously mentioned, we consider here a Distributed CSIT configuration in which each TX receives a different imperfect estimate of the multi-user channel\cite{dekerret2012_TIT}.  
			For sake of exposition, we consider that the CSI accuracy available at TX~$j$ is \emph{homogeneous} across RXs. %, i.e., that the accuracy is the same for both RXs' channels. 
			Note that  our results are not restricted by this assumption and they extend to the case with different accuracy for each RX. 
						
			It is known that,  in order to avoid the collapse of DoF in the Centralized CSIT setting, the CSIT error variance has to scale as~$P^{-\alpha}$, with $\alpha\! > 0$, \cite{Jindal2006,Davoodi2016_TIT_DoF}, where $\alpha$ is called the  \emph{CSIT scaling coefficient}. Based on that result, we extend the model to the distributed setting by assuming that the  error variance at TX~$j$ scales as $P^{-\alpha\expj}$, with $\alpha\expj>0$ and $\alpha\expo \neq \alpha\expt$.

			Specifically, we consider that RX~$i$ feeds back to TX~$j$ a quantized version of the normalized vector  $\thv_i \triangleq \frac{\hv_i}{\norm{\hv_i}}\in \mathbb{C}^2$ using  $B^{(j)}$~bits, denoted as $\bhh_i^{(j)}$. 
			We assume that RX~$i$ uses random vector quantization codebooks of~$2^{B^{(j)}}$ codewords \cite{Jindal2006}, such that the codewords $\hat{\bh}_i^{(j)}$ are unit-norm vectors uniformly distributed on the $2$-dimensional complex unit sphere. 	
			After receiving the feedback from both RXs, TX~$j$ obtains a multi-user channel estimate~$\hat{\bH}^{(j)}=[\bhh_1^{(j)},\bhh^{(j)}_2]^{\He} \in \mathbb{C}^{2\times 2}$. 
			In order to avoid degenerated conditions, we assume that the codebooks of different RXs do not share any codeword.
			%Our findings hold for any CSI codebooks with quantization error scaling as $2^{-B_i^{(j)}}$, but we consider here only random vector quantization for simplicity.
			%Furthermore, we let the number of quantization bits grow with  $P$ as
			Moreover, we let the number of quantization bits grow linearly with  $\log_2(P)$ as
				\begin{equation}
						B^{(j)} = \alpha^{(j)}\log_2(P).				\label{eq:SM_6}
				\end{equation}
			This implies that the CSIT error variance at TX~$j$ scales as $P^{-\alpha\expj}$  (since $P^{-\alpha\expj} = 2^{-B\expj}$\cite{Jindal2006}). 				
			%This implies that $P^{\alpha^{(j)}} = 2^{B^{(j)}}$. 
			%Under such feedback condition, it is known that the DoF of our setting is equal to $1 + \min(1, \max(\alpha^{(1)},\alpha^{(2)}))$ \cite{dekerret2012_TIT}, whereas this DoF collapses if the number of bits does not scale linearly with $\log_2(P)$ \cite{Jindal2006, Davoodi2016_TIT_DoF}.
			%The scalar $\alpha^{(j)}$ is called the \emph{CSIT scaling coefficient} at TX~$j$.  
			Under such feedback condition, it is known that the multiplexing gain (DoF) of our setting is equal to $1 + \min(\max(\alpha^{(1)},\alpha^{(2)}),1)$ \cite{dekerret2012_TIT}, whereas this multiplexing gain collapses if the number of bits does not scale linearly with $\log_2(P)$ \cite{Jindal2006, Davoodi2016_TIT_DoF}. 
			%Its value can be restricted to $[0,1]$, where $\alpha^{(j)}=0$ is generally seen to correspond to a CSIT being useless at high-SNR regime ---since it does not scale with $P$--- whereas $\alpha^{(j)}=1$ is sufficient to achieve the full multiplexing gain\cite{Jindal2006,Davoodi2016_TIT_DoF}. 
			We assume that both $\alpha^{(j)}$ are strictly positive.
      Given that one TX has the same CSIT quality ($\alpha^{(j)}$) for all the links, we can order them w.l.o.g. such that
				\begin{align}
						1 \geq \alpha^{(1)} \geq \alpha^{(2)} > 0. \label{eq:distributed_alpha_order} 
				\end{align}
			%where the value of $\alpha^{(2)}$ is strictly positive because it is required that the CSIT quality scales with the SNR to avoid the collapse of the DoF\cite{Jindal2006, Davoodi2016_TIT_DoF}. %, and therefore finite precision CSIT scenarios are precluded from this work. 
			%Importantly, %we consider that due to delay or backhaul constraints no additional communications are allowed between the TXs, such that 
			%the transmit coefficients at TX~$j$ are designed \emph{exclusively} on the basis of its corresponding $\hat{\bH}^{(j)}$ and the channel statistics. %, without any additional communication to the other TX.
			The multi-user distributed CSIT configuration is represented through the multi-TX CSIT scaling vector $\bm{\alpha}\in \mathbb{R}^{2}$ defined as
				\begin{equation}
						\bm{\alpha}\triangleq \begin{bmatrix} \alpha^{(1)} \\ \alpha^{(2)}\end{bmatrix}.		\label{eq:alpha_vector}
				\end{equation}   							
			Importantly, we consider that due to delay constraints no additional communications are allowed between the TXs, such that the transmit coefficients at TX~$j$ are designed \emph{exclusively} on the basis of its corresponding $\hat{\bH}^{(j)}$ and the channel statistics, without any additional communication to the other TX.  
 
		%%%%%%%%%%%%%%%%%%%%%%%%%%%%%%%%%%%%%%%%%%%%%%%%%%%%%
		%%%%%%%%%%%%%%%%%%%%%%%%%%%%%%%%%%%%%%%%%%%%%%%%%%%%%
			\subsection{Figure-of-Merit}
		%%%%%%%%%%%%%%%%%%%%%%%%%%%%%%%%%%%%%%%%%%%%%%%%%%%%%It is defined as $R \triangleq R_1 + R_2$,
		%%%%%%%%%%%%%%%%%%%%%%%%%%%%%%%%%%%%%%%%%%%%%%%%%%%%%
			Our figure-of-merit is the expected sum rate over both the fading realizations and the random codebooks. Let us define the expected rate of RX~$i$ as $R_i \triangleq \ExpHC{r_i}$, where $r_i$ is the instantaneous rate of RX~$i$. In our setting, $r_i$ writes as
				\begin{equation}
						 r_i\triangleq\log_2\bigg( 1+\frac{\frac{P}{2}|\hv_i^{\He}\tv_i|^2}{1+\frac{P}{2}|\hv_i^{\He}\tv_{\bar{i}}|^2}\bigg), 			\label{eq:fig_of_merit}
				\end{equation} 
			where we have introduced the notation $\bar{i} \triangleq i\!\pmod{2} + 1$. Then, the expected sum rate is given by $R\triangleq R_1+R_2$.
			%, and the expectation is done over both the fading realizations and the random codebooks.

		%%%%%%%%%%%%%%%%%%%%%%%%%%%%%%%%%%%%%%%%%%%%%%%%%%%%%
		%%%%%%%%%%%%%%%%%%%%%%%%%%%%%%%%%%%%%%%%%%%%%%%%%%%%%
			\subsection{Centralized Zero-Forcing Precoding}\label{se:sysmod_centr_zf}
		%%%%%%%%%%%%%%%%%%%%%%%%%%%%%%%%%%%%%%%%%%%%%%%%%%%%%
		%%%%%%%%%%%%%%%%%%%%%%%%%%%%%%%%%%%%%%%%%%%%%%%%%%%%%
			We restrict this work to ZF precoding schemes, which are known to achieve the optimal DoF in the centralized CSIT setting\cite{Jindal2006, Davoodi2016_TIT_DoF} and that allow for analytical tractability. 
			In this ``{ideal}'' centralized setting, all the TXs have access to the same channel estimate $\hat{\bH}$. 
			Similarly to the distributed CSIT case, we define $\bhh_i^{\He}$ as the shared estimate of the normalized vector channel towards RX~$i$, obtained with a feedback rate of $B=\alpha\log_2(P)$~bits. 
			Hence, in the centralized case $\bhh_i = {\bhh}_i^{(j)}$ $\forall j\in \{1,2\}$.   
			Let  $\vv^{\star}_i$ denote a unit-norm ZF precoder for RX~$i$, computed on the basis of the estimate~$\hat{\bH}$.  
			We can then write the centralized ZF precoding matrix  as $\bT^{\ZF}\triangleq\begin{bmatrix}\mu_1\vv_1^{\star}&\mu_2\vv_2^{\star}\end{bmatrix}$, where~$\mu_i\in\Rb$ is a parameter that ensures that the instantaneous power constraint $\norm{\tv_{\TX j}}\leq 1$  is fulfilled,    and which will be detailed later.
			From the  ZF precoding definition, $\vv^{\star}_i$ is a vector satisfying that %any unit-norm  vector satisfying 
				\eqm{
						%\qquad\qquad
						{\bhh}_{\bar{i}}^{\He}\vv_i^{\star}=0.	%\qquad \bar{i} = i\!\!\!\!\!\pmod{2} + 1. 
						\label{eq:zf_condition}
				}
			Given that %It follows from~\eqref{eq:zf_condition} that 
			multiplying the beamformer~$\vv^{\star}_i$ by a phase-shift~$e^{\imath\phi_i}$ does not impact the rate\cite{Wiesel2008},  
			%Thus, we can assume that the elements of the last row of the precoder belong to~$\Rb$. 
			%Consequently,  w.l.o.g. we select among all the possible 
			we can select w.l.o.g., among all the possible 
			$\vv_i^{\star}$, the vector $\vv_i^{} =  e^{-\imath\phi^v_i}[\hh_{\bar{i}2},\ -\hh_{\bar{i}1}]^\Trans$, where $\phi^v_i$ is the phase of the second coefficient $(\hh_{\bar{i}1})$. % ${{\vv_{i,2}^{\ZF}}} = -\hh_{\bar{i}1}$. 
			Thus, \vspace{-.5ex} %we have
				\begin{equation}
						\bT^{\ZF} =
							\underbrace{
								\begin{bmatrix} 
										\hh_{21}^{-1} \hh^{}_{22}    			& \hh_{11}^{-1}\hh^{}_{12} \\
										- 1 																			& -  1
								\end{bmatrix}\vspace{-1ex} 
							}_{\triangleq \bV^{\star}}
							\underbrace{							
								\begin{bmatrix} 
										\lambda^{\star}_1  & 0 \\
										0	 &  \lambda^\star_2 
								\end{bmatrix}			
							}_{\triangleq\bm{\Lambda}^{\star}}, 	\vspace{-.5ex} 					\label{eq:zf_split}
				\end{equation}
			where we have introduced the notation~$\lambda^{\star}_i \triangleq \mu_i{|{\vv_{i,2}^{}}|}$ From the unitary power constraint  it holds that $0\leq \lambda^\star_i \leq 1$. 
			%The term $\{\vv_i^{\ZF}\}_2$ denotes the second element of the precoding vector $\vv_i^{\ZF}$ and in this case it is given by $\{\vv_i^{\ZF}\}_2= \hh_{21}$. 
			Expression in~\eqref{eq:zf_split} is just a rewriting of the conventional ZF matrix used in the literature\cite{Jindal2006}, introduced to make the analogy with the distributed approach more explicit, such that we detach the interference-nulling part ($\bV^{\star}$) and the power control ($\bm{\Lambda}^{\star}$).

		%%%%%%%%%%%%%%%%%%%%%%%%%%%%%%%%%%%%%%%%%%%%%%%%%%%%%
		%%%%%%%%%%%%%%%%%%%%%%%%%%%%%%%%%%%%%%%%%%%%%%%%%%%%%
			\subsection{Instantaneous Power Control}
		%%%%%%%%%%%%%%%%%%%%%%%%%%%%%%%%%%%%%%%%%%%%%%%%%%%%%
		%%%%%%%%%%%%%%%%%%%%%%%%%%%%%%%%%%%%%%%%%%%%%%%%%%%%%							
			The power normalization strategy is performed by $\mu_i$ and follows any algorithm that belongs to a broad family of functions satisfying the per-TX instantaneous power constraint	$\norm{\tv_{\TX j}}\leq 1$, $\forall j\in\{1,2\}$. We recall that the term \emph{instantaneous} refers only to the precoding vector power. The transmit power satisfies an \emph{average} power constraint as it depends on the information symbols.
			Specifically, let $\lambda_i$ be the power-control value for RX~$i$'s symbols, such that $\lambda^{}_i \triangleq \mu_i{|{\vv_{i,2}^{}}|}$. 
			We model the power control as a function $\Lambda$  such that  $\forall i\in\{1,2\}$,
				\eqm{
						\lambda_i = \Lambda\LB\hat{\bH}, \bm{\alpha}, P,i\RB,
				}			
			where $\lambda^{}_i \in\Rb$. We assume that $\Lambda$ is $C^1$, i.e., all its partial derivatives exist and are continuous, and that its Jacobian Matrix $\bJ_{\Lambda}$ satisfies $\norm{\bJ_{\Lambda}}\leq M_\bJ  < \infty$. Moreover, the probability density function of $\Lambda_i$, denoted as~$f_{\Lambda_i}$ is bounded away from infinity such that
				\eqm{
						\max_{x}f_{\Lambda_i}(x) \leq f^{\max}_{\Lambda_i} < \infty. \label{eq:bound_pdf}
				}
			From the RVQ feedback assumption, $\hat{\bH}$ is distributed as $\tilde{\bH}$ and hence the marginal pdf $f_{\Lambda_i}(x)$ is the same for perfect, imperfect centralized and distributed CSIT.  To conclude, since the power control acts on the normalized precoder, the instantaneous power constraint per TX implies that
				\eqm{
						0 \leq \lambda^{}_i \leq 1. \label{eq:power_control_bound}
				}	
					%\begin{remark}
			%The constant $\eta_{\Lambda}> 0$ precludes degenerated cases and it can be arbitrarily small. Moreover, $\Lambda$ is a function of $P$ but it does not scale as it, since $M_{\Lambda}$ is independent of $P$. 
					%\end{remark}

	%%%%%%%%%%%%%%%%%%%%%%%%%%%%%%%%%%%%%%%%%%%%%%%%%%%%%
	%%%%%%%%%%%%%%%%%%%%%%%%%%%%%%%%%%%%%%%%%%%%%%%%%%%%
	%%%%%%%%%%%%%%%%%%%%%%%%%%%%%%%%%%%%%%%%%%%%%%%%%%%%%
		\section{Main Results}\label{se:Main_Results}
	%%%%%%%%%%%%%%%%%%%%%%%%%%%%%%%%%%%%%%%%%%%%%%%%%%%%%
	%%%%%%%%%%%%%%%%%%%%%%%%%%%%%%%%%%%%%%%%%%%%%%%%%%%%%
	%%%%%%%%%%%%%%%%%%%%%%%%%%%%%%%%%%%%%%%%%%%%%%%%%%%%%

		Although ZF precoding schemes as the one described in Section~\ref{se:sysmod_centr_zf} perform properly with centralized CSIT, their performance shrinks considerably on the distributed CSIT setting. This comes from the fact that the zero-forcing accuracy is proportional to the worst quality among the TXs ($\alpha\expt$ in our setting). Thus, conventional ZF does not achieve the centralized DoF.
		Furthermore, if TX~1 tries to estimate TX~2's CSI based on its own estimate it will incur in an estimation error proportional to $\alpha\expt$.
		
		The solution proposed in DoF-achieving schemes\cite{dekerret2012_TIT,Bazco2018_WCL,Bazco2018_TIT} --i.e., that TX~$2$ precodes with a vector independent of its instantaneous CSI-- also succumbs to the assumption of  instantaneous power constraint for the precoding vector ($\norm{\tv_{\TX j}}\leq 1$), since a less practical average power constraint was considered.		
		%Remarkably, considering an instantaneous power constraint, all the schemes presented in \cite{dekerret2012_TIT,Bazco2018_WCL,Bazco2018_TIT} do not achieve the optimal DoF. 
		The only scheme achieving the optimal DoF is obtained from \cite{dekerret2012_TIT} where the transmit power scales in ${P}/{\log(P)}$. This leads to a very inefficient power normalization, and hence to a very poor rate offset ($\Lc_\infty$).		
		
		We present a distributed precoding scheme, coined \emph{Hybrid Active-Passive ZF Precoding} (HAP), 
		%that eludes that TX~$2$'s accuracy affects the performance. 
		that precludes TX~$2$ from harming the performance. 
%Since the rate of centralized ZF is asymptotically achieved, and ZF is DoF-optimal in the centralized setting, Theorem~\ref{thm:optimality} also implies that we recover the DoF of the centralized setting. 
		The key for attaining such result is an asymmetric ZF scheme and the quantization of the power control, that allows the TXs to be \emph{consistent}. %  and {agree}.

		%Indeed, without quantization of the power values $\lambda^{(j)}_i$, it is easy to see from~\eqref{eq:hap_precoder_def} that the orthogonality is lost. 

		%%%%%%%%%%%%%%%%%%%%%%%%%%%%%%%%%%%%%%%%%%%%%%%%%%%%
		%%%%%%%%%%%%%%%%%%%%%%%%%%%%%%%%%%%%%%%%%%%%%%%%%%%%%
			%\subsection{Proposed Precoding Scheme: HAP}\label{subse:precoder_hap}
			\subsection{Proposed Hybrid Active-Passive ZF Precoding}\label{subse:precoder_hap}
		%%%%%%%%%%%%%%%%%%%%%%%%%%%%%%%%%%%%%%%%%%%%%%%%%%%%%
		%%%%%%%%%%%%%%%%%%%%%%%%%%%%%%%%%%%%%%%%%%%%%%%%%%%%%	
			%Let us  introduce the proposed distributed precoding scheme, coined  \emph{Hybrid Active-Passive ZF Precoding} (HAP), before presenting the main findings.
			Let $\Qc(\cdot)$~represent the output of an arbitrary quantizer $\Qc$ satisfying that $\Qc(x) \leq x$. 
				The HAP precoder, denoted by $\bT^{\HAP}\in\Cb^{2\times 2}$, is given by							
				The HAP precoder, denoted by $\bT^{\HAP}\in\Cb^{2\times 2}$, is given by		
					\eqm{
							\bT^{\HAP}\triangleq
								\begin{bmatrix} 
											 \big(\hh^{(1)}_{21}\big)^{-1}\hh^{(1)}_{22}   	&\big(\hh^{(1)}_{11}\big)^{-1}\hh^{(1)}_{12} \\
																	-	1 																					&	  -1
								\end{bmatrix}  
							\odot 
								\begin{bmatrix} 
										\mathcal{Q}(\lambda^{(1)}_1) 	& \mathcal{Q}(\lambda^{(1)}_2)\\
										\mathcal{Q}(\lambda^{(2)}_1) 	& \mathcal{Q}(\lambda^{(2)}_2) 
								\end{bmatrix} 				\label{eq:hap_precoder_def}
					}
				where $\odot$ denotes the Hadamard (element-wise) product and $\lambda^{(j)}_i$ is the distributed counterpart of $\lambda^\star_i$. 				
				%Note that, if we provide both TXs with $\hat{\bH}^{(j)}$ (the most accurate CSIT), then  $\lambda^{\star}_i = \lambda^{(1)}_i$. 
						%\begin{align}
								%$\lambda^{(j)}_i \triangleq {\mu_i^{(j)}|{\{\vv_i^{\ZF,(j)}\}_2}|}$, $\forall i,j\in\{1,2\}$,  %\label{eq:lambda_def}%\qquad \ell = i\!\!\!\!\pmod{2} + 1, \label{eq:lambda_def}
								%$\lambda^{(j)}_i \triangleq\Lambda_i\big(\hat{\bH}^{(j)}\big)$, $\forall i,j\in\{1,2\}$.  %\label{eq:lambda_def}%\qquad \ell = i\!\!\!\!\pmod{2} + 1, \label{eq:lambda_def}
						%\end{align}		
				%and $\vv_i^{\ZF,(j)}$ is the conventional ZF precoder using the imperfect channel knowledge~$\hat{\bH}^{(j)}$ available at TX~$j$. 
				We  observe that the first matrix is equal to the interference-nulling matrix~$\bV^{\star}$ in \eqref{eq:zf_split} based on the imperfect CSIT knowledge~$\hat{\bH}^{(1)}$, and hence it is independent of the CSI of TX~$2$. Conversely, the second matrix needs to be computed at both TXs, and thus it differs from the centralized power normalization matrix~$\bm{\Lambda}^{\star}$.
				The idea behind this separation is that the interference-nulling has to be extremely accurate, but it can be performed by a single TX, whereas the power normalization has to be done by both TXs, but it can be computed with a reduced precision, allowing the TXs to be consistent. 

				Since $\lambda^{(j)}_i\in [0,1$], we have that $\Qc(\lambda^{(j)}_i)\in[0, 1]$. 
				Moreover, we assume that it exists~$M_\Qc<\infty$ such that
					\eqm{
							\abs{\ExpHCq{\log_2\big({\Qc(x)}\big)}} \leq M_\Qc, \label{eq:cond_bounded} \tag{P0}
					}
				which is a technical assumption that is satisfied by any non-degenerate quantizer. 
				The role of $\Qc$ is to trade-off the accuracy of the power control with the consistency of the decision at the TXs, as the ZF orthogonality of~\eqref{eq:zf_condition} is preserved only if both TXs obtain the same quantization value --if $\Qc\big(\lambda^{(1)}_i\big)\!=\!\Qc\big(\lambda^{(2)}_i\big)\!$--.  
				%Remarkably, the HAP precoder does not set a specific $\Qc$ and it can be chosen among a broad family of quantizers satisfying two asymptotic conditions that are presented in the following.			
				%Specifically, $\Qc$ can be any quantizer such that $\Qc(x) \leq x$. 
				In order to emphasize the relevance of the quantizer, we define $\Omega$ as the set of estimates $(\hat{\bH}^{(1)}\!\!,\hat{\bH}^{(2)})$ that ensure that the ZF orthogonality is not violated, excluding degenerate cases, i.e.,	
					\eqm{
							&\Omega \triangleq \left \{(\hat{\bH}^{(1)},\hat{\bH}^{(2)})\big| \ \forall i\in \{1,2\}\quad   \Qc\big(\lambda^{(1)}_i\big)=\Qc\big(\lambda^{(2)}_i\big) \in\Rb^+\,\right\}.		\label{eq:omega_def} 
					}												
				In simple words, $\Omega$ encloses the cases when the TXs agree on the power normalization coefficients for both RXs and they are strictly positive. We further denote the complementary event of $\Omega$ as $\Omega^\setcomp$ (the \emph{inconsistent} cases). %, such that $\Pr(\Omega\cup\Omega^\setcomp)=1$ and $\Pr(\Omega\cap\Omega^\setcomp)=0$.
				We proceed by introducing two important properties for the quantizers.
				%present in the following a type of precoder that will be useful for showing our main results.
					\begin{definition}[Asymptotically Accurate Quantizers] 
							%Let $\alpha^{(1)}\geq\alpha^{(2)}$. 
							A quantizer $\Qc$ is said to be \emph{asymptotically accurate} if 
								\begin{align}
											\lim_{P\rightarrow \infty}\Qc(\lambda^{(j)}_i)=\lambda_i^{(1)}\quad\text{a.s.}\quad \forall i,j\in\{1,2\},  \label{eq:cond_convergence}			\tag{P1}
									\end{align}				
							where a.s. stands for \emph{almost surely}.
					\end{definition}		
					\begin{definition}[Asymptotically Consistent Quantizers] 
							%Let $\alpha^{(1)}\geq\alpha^{(2)}$. 
							A quantizer $\Qc$ is said to be  \emph{asymptotically consistent} if 
								\eqm{
										 \Pr\left(\Omega^\setcomp\right)=o\LB\frac{1}{\log_2(P)} \RB\!\!\, .  \label{eq:cond_agreement} \tag{P2}
								}
							%where $\Pr(A)$ stands for the probability of the event $A$. 	
					\end{definition}										
				%A quantizer that is both \emph{asymptotically accurate}  and \emph{asymptotically consistent} is said to be an \emph{Asymptotically Optimal Quantizer}.
				Property~\eqref{eq:cond_agreement} implies that  \emph{inconsistent precoding} events are negligible in terms of asymptotic rate. 				
				We exhibit in the following lemma one particular quantizer satisfying properties~\eqref{eq:cond_convergence}-\eqref{eq:cond_agreement}. %, which is introduced in the following lemma. 
				Optimizing further this quantizer is crucial to good performance at finite SNR and its optimization is an ongoing research topic.  
					\begin{lemma}\label{lem:uniform_quantizer}
							Let $\Qc_u$ be a uniform quantizer in the interval~$[0,1]$ with a step size of $\Pb^{\frac{-\alpha^{(2)}}{2}}$,	 such that
									\eqm{
											\Qc_u(x) \triangleq {\Pb^{\frac{-\alpha^{(2)}}{2}}}\big\lfloor \Pb^{\frac{\alpha^{(2)}}{2}}x\big\rfloor. \label{eq:agree_proof_0ab}
									}
							Then, $\Qc_u$ satisfies properties~\eqref{eq:cond_bounded},~\eqref{eq:cond_convergence} and~\eqref{eq:cond_agreement}. %$\abs{\ExpHCq{\log_2\big({\Qc_u(x)}\big)}} \leq M$
					\end{lemma}
					\begin{IEEEproof}		
							The proof is relegated to Appendix~\ref{app:uniform_quantizer}. % extended version due to space constraints~\cite{Bazco2019_ISIT_extended}.  
					\end{IEEEproof}

		%%%%%%%%%%%%%%%%%%%%%%%%%%%%%%%%%%%%%%%%%%%%%%%%%%%%
		%%%%%%%%%%%%%%%%%%%%%%%%%%%%%%%%%%%%%%%%%%%%%%%%%%%%%
			\subsection{Main Results}\label{subse:Main_Results}
		%%%%%%%%%%%%%%%%%%%%%%%%%%%%%%%%%%%%%%%%%%%%%%%%%%%%%
		%%%%%%%%%%%%%%%%%%%%%%%%%%%%%%%%%%%%%%%%%%%%%%%%%%%%%	
						
				Let us denote by $R^{\HAP}(\alpha^{(1)},\alpha^{(2)})$ the expected sum rate achieved using HAP precoding in the Distributed CSIT setting with CSIT scaling quality~($\alpha^{(1)},\alpha^{(2)}$). 
				Similarly, we denote as~$R^{\ZF}(\alpha^{(1)})$ the expected sum rate attained by the centralized ZF precoder of Section~\ref{se:sysmod_centr_zf} on the basis of the estimate~$\hat{\bH}^{(1)}$. % with scaling coefficient~$\alpha$. 
				Accordingly, the  rate gap between those settings  is defined as   
					\eqm{
							{\DeltaR} \triangleq {R^{\ZF}(\alpha^{(1)})}	-		 {R^{\HAP}(\alpha^{(1)},\alpha^{(2)})}.			\label{eq:def_deltaR}
					}
				%and its instantaneous counterpart denoted as ${\Delta r}$, such that ${\Delta r} \triangleq {r_1^{\ZF}(\alpha^{(1)})}	-		 {r_1^{\HAP}(\alpha^{(1)},\alpha^{(2)})}$ and $\DeltaR = \ExpHC{\Delta r}$.  
				%We further denote the conditional expectation given an event $A$ as $\Eb_{|A}$. % or $R_{|A}$.  		
				%%%%%%%%%%%%%%%%%%%%%%%%%%%%%%%%%%%%%%%%%%%%%%%%%%%%%%
				We can now state our main results. 
					\begin{theorem}\label{thm:bound_gap}
						The rate gap of ZF precoding with Distributed CSIT  is upper bounded by
						%it holds that %linear precoding asymptotically achieves the same sum rate as in the centralized scenario with  CSI accuracy $\alpha^{(1)}$. Specifically, the Hybrid Active-Passive ZF Precoding attains such optimal performance, i.e.,
							\begin{align}
									 {\DeltaR} &\leq 2\ExpHCo{\log_2\LB {\Gamma_1}\RB} + \Pr\LB\Omega^\setcomp\RB {R^{\ZF}_{|\Omega^\setcomp}(\alpha^{(1)})},				  \label{eq:Main_Result_1} %{R^{\ZF}(\alpha^{(1)})}
							\end{align}
						where $\Omega$ is defined in~\eqref{eq:omega_def}, $\Gamma_1$ is defined as % \triangleq \big|{\lambda^{(1)}_1}/{\Qc(\lambda^{(1)}_1)}\big|^2$, and . 
							\eqm{
								\Gamma_1 \triangleq \bigg|\frac{\lambda^{(1)}_1}{\Qc(\lambda^{(1)}_1)}\bigg|^2,
							}	
						and it holds that ${R^{\ZF}_{|\Omega^\setcomp}(\alpha^{(1)})}\leq 2\log_2\LB 1 +  P\RB$.
					\end{theorem} 
				%%%%%%%%%%%%%%%%%%%%%%%%%%%%%%%%%%%%%%%%%%%%%%%%%%%%%%
				%This bound holds for any quantizer $\Qc$ with $\Qc(x)\leq x$, and for this reason~\eqref{eq:Main_Result_1} has a general expression  that depends on the set $\Omega$ and thus on the $\Qc$ selected. 
				The proof is detailed in Section~\ref{se:Proof}.
				This bound  depends on the set $\Omega$ and thus on the quantizer  selected. 
				Intuitively, a \emph{good} quantizer has to  ensure a high probability of agreement, so as to make $\Pr\LB\Omega^\setcomp\RB$ small. This can be done by enlarging the quantization step, what will make the first term bigger, as ${\Qc(\lambda^{(1)}_1)}$ needs to be as close to ${\lambda^{(1)}_1}$ as possible. This shows why finding the optimal quantizer is a challenging research topic. Nevertheless, there exists a family of quantizers that behave asymptotically optimal, as stated in the following theorem.
				
				%%%%%%%%%%%%%%%%%%%%%%%%%%%%%%%%%%%%%%%%%%%%%%%%%%%%%%
					\begin{theorem}\label{thm:optimality}
						Let $\Qc$ be an arbitrary quantizer satisfying \eqref{eq:cond_bounded},~\eqref{eq:cond_convergence} and~\eqref{eq:cond_agreement}.
						%the uniform quantizer defined in Lemma~\ref{lem:uniform_quantizer}. %, %a quantizer that satisfies  properties~\eqref{eq:cond_convergence} and~\eqref{eq:cond_agreement}. 
						Then, taking the limit in Theorem~\ref{thm:bound_gap} %of~\eqref{eq:Main_Result_1} when $P\rightarrow\infty$ 
						yields 
						%it holds that
							\begin{equation}
									\lim_{P\rightarrow \infty}\DeltaR \leq 0.				\label{eq:Main_Result_2}
							\end{equation}
					\end{theorem} 
						\begin{IEEEproof}
								The proof follows from Theorem~\ref{thm:bound_gap}. First, note that the sum rate ${R_{|\Omega^\setcomp}^{\ZF}(\alpha^{(1)})}$  is trivially bounded by twice the interference-free single-user rate to obtain %that 
									\eqm{
											{R_{|\Omega^\setcomp}^{\ZF}(\alpha^{(1)})} 
													& \leq \sum_{i=1}^2 \log_2\LB 1 + \frac{P}{2}\,\ExpHC{\norm{\hv_i}^2}\RB \\
													& =    2\log_2\LB 1 + P\RB,  \label{eq:proof_bound_omegac_2}
									}
								what together with property~\eqref{eq:cond_agreement} implies that
									\eqm{
											\Pr\LB\Omega^\setcomp\RB {R_{|\Omega^\setcomp}^{\ZF}(\alpha^{(1)})}  = o(1). \label{eq:proof_omega_c_vanish}
											%\Pr\LB\Omega^\setcomp\RB {R^{\ZF}(\alpha^{(1)})} = o(1), \label{eq:proof_omega_c_vanish}
									}
								%what together with property~\eqref{eq:cond_agreement} implies that
									%\eqm{
											%\Pr\LB\Omega^\setcomp\RB {R_{|\Omega^\setcomp}^{\ZF}(\alpha^{(1)})}  = o(1). \label{eq:proof_omega_c_vanish}
											%%\Pr\LB\Omega^\setcomp\RB {R^{\ZF}(\alpha^{(1)})} = o(1), \label{eq:proof_omega_c_vanish}
									%} 
								%since we can bound the centralized rate by the perfect-CSIT, single-user setting and thus ${R^{\ZF}(\alpha^{(1)})} \leq 2\log_2(1+2P)$.
								Consequently, to conclude the proof it only remains to show that~$\limpf \ExpHCo{\log_2(\Gamma_1)} =0$. From the definition of $\Gamma_1$, it holds that
									\eqm{
											&\!\ExpHCo{\log_2\LB\Gamma_1 \RB} 
												%= \ExpHCo{\log_2\bigg(\frac{\lambda^{(1)}_1}{\Qc(\lambda^{(1)}_1)} \bigg)}\\
												%&\hspace{5ex}
												=\! \ExpHCo{\log_2\!\big({\lambda^{(1)}_1}\big)\!} - \ExpHCo{\!\log_2\!\big({\!\Qc(\lambda^{(1)}_1)} \big)\!}. \label{eq:proof_theo_limit_2}
									}								
								Note that, for any variable $x$ such that $0\leq x\leq 1$, and for any two events $A,B,$ such that $0 <\Pr(B\mid A) < 1$, it holds that
									\eqm{
										\Eb_{\mid A}[\log_2(x)] &= \Pr(B\mid A)\Eb_{\mid A\cap B}[\log_2(x)] + \Pr(B^\setcomp\mid A)\Eb_{\mid A\cap B^\setcomp}[\log_2(x)].
									}
								Since $0\leq x\leq 1$, $\Eb_{\mid A\cap B^\setcomp}[\log_2(x)]\leq 0$ and hence
									\eqm{
										\Eb_{\mid A\cap B}[\log_2(x)] 
												%& = \frac{1}{\Pr(B\mid A)}\LB\Eb_{\mid A}[\log_2(x)] - \Pr(B^\setcomp\mid A)\Eb_{\mid A\cap B^\setcomp}[\log_2(x)]\RB\\
												& \geq \frac{1}{\Pr(B\mid A)}\Eb_{\mid A}[\log_2(x)]. \label{eq:bound_max_int} 
									}					
								%what comes from the fact that $\Eb_{\mid A\cap B^\setcomp}[\log_2(x)]\leq 0$ as $0\leq x\leq 1$. 								
								Therefore, if $\Eb_{\mid A}[\log_2(x)]$ exists, also $\Eb_{\mid A\cap B}[\log_2(x)] $ exists and it is bounded below by~\eqref{eq:bound_max_int} and above by $0$. 
								Let  $A$ and $B$ be   $A= \{\Qc(\lambda^{(1)}_i)>0,\forall i\}$ and $B = \big\{\Qc\big(\lambda^{(1)}_i\big)\!=\!\Qc\big(\lambda^{(2)}_i\big), \forall i\}$. Thus, $\Omega = A\cap B$. 
								It follows from~\eqref{eq:bound_max_int} and~\eqref{eq:cond_bounded} that
									\eqm{
											\Eb_{\mid\Omega}[\log_2\big({\Qc(\lambda^{(1)}_1)} \big)] \geq  -\frac{\Pr( \Qc(\lambda^{(1)}_i)>0,\forall i)}{\Pr(\Omega)}M_\Qc,
									}
								where we have applied the fact that $\Pr(B|A)=\frac{\Pr(A\cap B)}{\Pr(A)}$. Hence, $\Eb_{\mid\Omega}[\log_2({\Qc(\lambda^{(1)}_1)} )] $ is bounded. 
								The same result follows for~$\Eb_{\mid\Omega}[\log_2\big({\lambda^{(1)}_1} \big)]$ from the bounded density assumption of~\eqref{eq:bound_pdf}.
								%The same result follows for~$\Eb_{\mid\Omega}[\log_2\big({\Qc_u(\lambda^{(1)}_1)} \big)]$ from property~\eqref{eq:cond_bounded}.
								%It follows from the bounded density assumption of~\eqref{eq:bound_pdf}  that~$\Eb_{\mid\Omega}[\log_2\big({\lambda^{(1)}_1} \big)]$ exists and it is bounded. 
								Moreover, from the continuity of the~$\log$ function and~\eqref{eq:cond_convergence}, $\log_2({\Qc(\lambda^{(1)}_1)})$ converges a.s. to $\log_2({\lambda^{(1)}_1})$. From all these facts,  
								we can apply  Lebesgue's Dominated Convergence Theorem\cite[Theorem~16.4]{Billingsley1995} to interchange expectation and limit and show that
									\eqm{
											\limpf \ExpHCo{\log_2\LB{\Qc(\lambda^{(1)}_1)} \RB}=  \ExpHCo{\log_2\LB{\lambda^{(1)}_1} \RB},
									}										
								%and then from~\eqref{eq:proof_theo_limit_2} we obtain that 
									%\eqm{
											%\limpf \ExpHCo{\log_2\LB\Gamma_1 \RB} =0. \label{eq:proof_omega_vanish}
									%}										
								and thus 
									%\eqm{
											%\limpf \ExpHCo{\log_2\bigg(\frac{\lambda^{(1)}_1}{\Qc(\lambda^{(1)}_1)} \bigg)} =0. \label{eq:proof_omega_vanish}
											$\limpf \ExpHCo{\log_2(\Gamma_1)} =0$, %\label{eq:proof_omega_vanish}
									%}
								%This result, together with~\eqref{eq:proof_omega_c_vanish}, implies that
									%\begin{align}
											 %\limpf  \ExpHCo{\log_2\LB\Gamma_1 \RB} + \Pr\LB\Omega^\setcomp\RB  {R_{|\Omega^\setcomp}^{\ZF}(\alpha^{(1)})}	= 0,	\label{eq:proof_theo_limit_5}
											 %%\limpf \Pr\LB\Omega\RB \ExpHCo{\log_2\LB\Gamma_1 \RB} + \Pr\LB\Omega^\setcomp\RB {R^{\ZF}(\alpha^{(1)})}	= 0	\label{eq:proof_theo_limit_5}
									%\end{align}						
								which concludes the proof. 								
					\end{IEEEproof}		
				%%%%%%%%%%%%%%%%%%%%%%%%%%%%%%%%%%%%%%%%%%%%%%%%%%%%%%
					\begin{corollary}[Rate Offset with HAP precoder]
							It holds from Theorem~\ref{thm:optimality} that the rate offset $\Lc_\infty\!$ --defined in~\eqref{eq:def_affine}-- of ZF with distributed CSIT is the same as for the genie-aided centralized setting, whose rate offset  was shown in~\cite{Jindal2006} to be constant with respect to Perfect CSIT ZF (and thus with respect to the capacity-achieving Dirty Paper Coding) for $\alpha=1$.  	
							Specifically, for a constant $b$, if $B^{(1)}=\log_2(P) - \log_2(b)$, then the rate offset with respect to Perfect CSIT ZF is given by $\log_2(b)$\cite{Jindal2006}. 
					\end{corollary}					
						%\paragraph*{Rate Offset with HAP precoder} It follows from Theorem~\ref{thm:optimality} that the rate offset $\Lc_\infty$ ---defined in~\eqref{eq:def_affine}--- of ZF in the distributed CSIT setting is the same as for the genie-aided centralized setting, whose rate offset  was shown in~\cite{Jindal2006} to be constant with respect to Perfect CSIT ZF (and thus with respect to the capacity-achieving Dirty Paper Coding\cite{Lee2007}) for $\alpha=1$. Specifically, for a constant $b$, if $B^{(1)}=\log_2(P) - \log_2(b)$, then the rate offset with respect to Perfect CSIT ZF is given by $\log_2(b)$\cite{Jindal2006}.   
																 %linear precoding asymptotically achieves the same sum rate as in the centralized scenario with  CSI accuracy $\alpha^{(1)}$. Specifically, the Hybrid Active-Passive ZF Precoding attains such optimal performance, i.e.,
																
				%%%%%%%%%%%%%%%%%%%%%%%%%%%%%%%%%%%%%%%%%%%%%%%%%%%%%%		
														
				%Since the rate of centralized ZF is asymptotically achieved, and ZF is DoF-optimal in the centralized setting, Theorem~\ref{thm:optimality} also implies that we recover the DoF of the centralized setting. 
				The key for attaining such surprising performance is the trade-off between \emph{consistency} and \emph{accuracy} that is ruled by the quantizer.   
				Indeed, without quantization of the power values $\lambda^{(j)}_i$, it is easy to see from~\eqref{eq:hap_precoder_def} that the orthogonality is lost. 
				%Conversely, using a quantizer that satisfies properties~\eqref{eq:cond_convergence} and~\eqref{eq:cond_agreement} enables us to reach a high probability of agreement such that the DoF is not reduced --property~\eqref{eq:cond_agreement}--, whereas asymptotically reaching the perfect normalization such that the power constraint is attained with equality --property~\eqref{eq:cond_convergence}--. % and there is asymptotically no rate loss from it. 
			%
				Interestingly, Lemma~\ref{lem:uniform_quantizer} illustrates that simple quantizers --as the uniform one-- satisfy the sufficient conditions of convergence if we select the correct number of quantization levels. Moreover, since this quantizer is applied locally and no information exchange is done, the granularity of the quantizer does not increase the complexity of the scheme. 
				
				Let us consider that there is agreement between the TXs, i.e., that $\Qc(\lambda^{(1)}_i)=\Qc(\lambda^{(2)}_i),\forall i\in \{1,2\}$, such that we can define
					\eqm{
							\quad \lambda^{\Qc}_i \triangleq \Qc\big(\lambda^{(j)}_i\big), \qquad\quad  \forall i,j\in\{1,2\}, \label{eq:lambda_q_def}
					}
				what~\eqref{eq:cond_agreement} ensures that occurs with a probability high enough such that the disagreement is asymptotically negligible. In this case %the two rows of the second matrix in \eqref{eq:Main_Result_4} are equal and 
				we can  rewrite~\eqref{eq:hap_precoder_def} %the element-wise product 
				with a conventional matrix multiplication to get
					\begin{equation}
							\bT^{\HAP}\!\triangleq \!
								\begin{bmatrix} 
											 \big(\hh^{(1)}_{21}\big)^{\!-1}\hh^{(1)}_{22}   	& \big(\hh^{(1)}_{11}\big)^{\!-1}\hh^{(1)}_{12} \\
																	-	1 																					&	  -1
								\end{bmatrix}							
							\begin{bmatrix} 
										 \lambda^{\Qc}_1 	&  0\\
										0 						&  \lambda^{\Qc}_2 
							\end{bmatrix}\!. 				\label{eq:hap_split}
					\end{equation} 
				It becomes then clear that the orthogonality (i.e., the interference attenuation) is ensured by the first matrix in \eqref{eq:hap_split} while the second diagonal matrix is only used to satisfy the power constraint. %for power control. 
				%It is this division between the orthogonality and the power control that is key to the proof of the theorem,  as the orthogonality has to be very precise ---in the order of $\bar{P}^{-\alpha^{(1)}}$--- whereas the requirements for the power  accuracy is much weaker.
				Regarding the quantizer $\Qc$, note that letting $\Qc$ having a single quantization point leads to a statistical power control, whereas letting $\Qc$ have infinite points leads to the unquantized version. In both cases, part of the DoF is lost. 
				%In order to show that properties~\eqref{eq:cond_convergence} and~\eqref{eq:cond_agreement} are easily met, we will present a simple quantizer satisfying them in Section~\ref{se:example}.
			
				%The second important limitation comes from the average power constraint considered in \cite{dekerret2012_TIT,Bazco2018_WCL,Bazco2018_TIT}. Such a constraint is not practical and is not in line with conventional assumptions. 			
				%The assumption of instantaneous power constraint stands in contrast with the previous works on the DoF of the Distributed Network MIMO, where a less practical average power constraint was considered \cite{dekerret2012_TIT,Bazco2018_WCL,Bazco2018_TIT}. 
				%Remarkably, considering an instantaneous power constraint, all the schemes presented in \cite{dekerret2012_TIT,Bazco2018_WCL,Bazco2018_TIT} do not achieve the optimal DoF. The only scheme achieving the optimal DoF is obtained from \cite{dekerret2012_TIT} where the transmit power scales in ${P}/{\log(P)}$ for a maximum instantaneous power of $P$. This leads to a very inefficient power normalization, and hence to a very poor rate offset.

	%%%%%%%%%%%%%%%%%%%%%%%%%%%%%%%%%%%%%%%%%%%%%%%%%%%%%
	%%%%%%%%%%%%%%%%%%%%%%%%%%%%%%%%%%%%%%%%%%%%%%%%%%%%%
	%%%%%%%%%%%%%%%%%%%%%%%%%%%%%%%%%%%%%%%%%%%%%%%%%%%%%
		\section{Proof of Theorem~\ref{thm:bound_gap}}\label{se:Proof}
	%%%%%%%%%%%%%%%%%%%%%%%%%%%%%%%%%%%%%%%%%%%%%%%%%%%%%
	%%%%%%%%%%%%%%%%%%%%%%%%%%%%%%%%%%%%%%%%%%%%%%%%%%%%%
	%%%%%%%%%%%%%%%%%%%%%%%%%%%%%%%%%%%%%%%%%%%%%%%%%%%%%		
		We consider  w.l.o.g.  the rate difference at RX~$1$, denoted as ${\DeltaR_1}$, since the proof for RX~$2$ is obtained after switching the indexes of the RXs. 	
		%Its inst%antaneous counterpart ${\Delta r_1}$ is defined similarly, such that $\DeltaR_1 = \ExpHC{\Delta r_1}$.  Then, we can write 
		${\DeltaR_1}$ can be split as
				\begin{equation}
						\DeltaR_1 =\Pr\LB\Omega\RB \DeltaR_{1|\Omega} +\Pr\LB\Omega^\setcomp\RB \DeltaR_{1|\Omega^\setcomp}.			\label{eq:proof_3b}
				\end{equation}
		%In order to  bound the absolute value of $\DeltaR$, we need to lower and upper bound $\DeltaR$. 
		%In the following, %we upper-bound $\DeltaR$ by bounding 
		%we bound separately the two terms of the right hand size of~\eqref{eq:proof_3b}. %The lower bound follows the same approach and it is detailed in the extended version~\cite{Bazco2019_ISIT_extended}. 
			%%
%
		%%%%%%%%%%%%%%%%%%%%%%%%%%%%%%%%%%%%%%%%%%%%%%%%%%%%%
		%%%%%%%%%%%%%%%%%%%%%%%%%%%%%%%%%%%%%%%%%%%%%%%%%%%%%
			%\subsection{Rate-offset conditioned on $\Omega$ (consistent precoding)}\label{subse:omega_condition}
		%%%%%%%%%%%%%%%%%%%%%%%%%%%%%%%%%%%%%%%%%%%%%%%%%%%%%				
		%%%%%%%%%%%%%%%%%%%%%%%%%%%%%%%%%%%%%%%%%%%%%%%%%%%%%
						%
			First, we focus on $\DeltaR_{1|\Omega}$, which encloses the \emph{consistent precoding} cases. 
			Conditioned on $\Omega$ it holds that $\Qc\big(\lambda^{(1)}_i\big)=\Qc\big(\lambda^{(2)}_i\big)$, $\forall i\in\{1,2\}$, and hence we can use the notation $\lambda^{\Qc}_i$ introduced in~\eqref{eq:lambda_q_def}. 
			%both TXs obtain the same value, we can denote the matched parameter for RX~$i$ as %$\lambda^{\Qc}_i = \Qc_{\alpha_^{(2)}}\big(\lambda^{(j)}_i\big)$. %, for any $j\in\{1,2\}$.
				%\eqm{
						%\qquad\quad \lambda^{\Qc}_i \triangleq \Qc\big(\lambda^{(j)}_i\big), \qquad\quad  \forall j\in\{1,2\}.
				%}
			Moreover, it can be observed from \eqref{eq:zf_split} and \eqref{eq:hap_split} that, conditioned on $\Omega$, the HAP precoder satisfies %the $\HAP$ precoder satisfies
				\eqm{
						\qquad\quad\tv^{\HAP}_i = \frac{\lambda^{\Qc}_i}{\lambda^{\star}_i}\tv^{\ZF}_i,\qquad\quad \forall i\in\{1,2\}.  \label{eq:proof_bounded_2}
				}
			%what follows from \eqref{eq:zf_split} and \eqref{eq:hap_split}. 
			Since we assume that in the centralized ZF setting both TXs share the channel estimate of TX~1 ($\hat{\bH}^{(1)}$), we have that $\lambda^{\star}_i = \lambda^{(1)}_i$. 
			Given that $\Qc(x)\leq x$, it follows that ${\lambda^{\Qc}_i}/{\lambda^{\star}_i} \leq 1$, $\forall i\in\{1,2\}$. 			
			Let us recall that $\Gamma_i$ is defined as %For sake of readability, we introduce the notation
				\eqm{
						\Gamma_i \triangleq \bigg|\frac{\lambda^{(1)}_i}{\lambda^{\Qc}_i}\bigg|^2, \label{eq:def_lambda_ratio}
				}
			which satisfies then that $\Gamma_i \geq 1$ $\forall i\in\{1,2\}$.
			%and $\lambda^{\Qc}_i\leq\lambda^{\star}_i$
			Thus, conditioned on $\Omega$ we can write that the SINR obtained through HAP precoding satisfies
				\eqm{
						%\ExpHCo{\log_2\LB1 + \frac{P\abs{\hv_1^{\He}\tv^{\HAP}_1}^2}{1 + P\abs{\hv_1^{\He}\tv^{\HAP}_2}^2}\RB} \\ %\hspace{13ex}						
						{1 + \frac{\frac{P}{2}\abs{\hv_1^{\He}\tv^{\HAP}_1}^2}{1 + \frac{P}{2}\abs{\hv_1^{\He}\tv^{\HAP}_2}^2}} 
								& = 1 +\frac{\frac{1}{\Gamma_1} \frac{P}{2}\abs{\hv_1^{\He}\tv^{\ZF}_1}^2}{1 + \frac{1}{\Gamma_2} \frac{P}{2}\abs{\hv_1^{\He}\tv^{\ZF}_2}^2} \label{eq:proof_bound_omega_1a} \\ %\hspace{13ex}						
								%& \geq 1 +\frac{ \frac{1}{\Gamma_1} P\abs{\hv_1^{\He}\tv^{\ZF}_1}^2}{1 +  P\abs{\hv_1^{\He}\tv^{\ZF}_2}^2} \label{eq:proof_bound_omega_1b}\\ %\hspace{13ex}						
								& \geq \frac{1}{\Gamma_1} \bigg(  1 +\frac{ \frac{P}{2}\abs{\hv_1^{\He}\tv^{\ZF}_1}^2}{1 +  \frac{P}{2}\abs{\hv_1^{\He}\tv^{\ZF}_2}^2}\bigg), \label{eq:proof_bound_omega_1c} %\hspace{13ex}						
				}
			%where~\eqref{eq:proof_bound_omega_1} comes from the fact that, given $\Omega$,  $\Lambda_i \leq 1$ $\forall i\in\{1,2\}$. 
			where~\eqref{eq:proof_bound_omega_1a} follows from~\eqref{eq:proof_bounded_2}-\eqref{eq:def_lambda_ratio} whereas~\eqref{eq:proof_bound_omega_1c} comes from the fact that $1/\Gamma_i \leq 1$ $\forall i$. %\eqref{eq:proof_bound_omega_1b} and
			%Furthermore, it holds that
				%\eqm{
						%{R_{1|\Omega}^{\ZF}(\alpha^{(1)})} = \ExpHCo{\log_2\LB  1 +\frac{ P\abs{\hv_1^{\He}\tv^{\ZF}_1}^2}{1 +  P\abs{\hv_1^{\He}\tv^{\ZF}_2}^2} \RB}. \label{eq:proof_bound_omega_2}
				%}
			%From~\eqref{eq:proof_bound_omega_1c},~\eqref{eq:proof_bound_omega_2} and  the fact that $-\log_2\LB\Gamma_1 \RB = \log_2\LB1/\Gamma_1 \RB$, we obtain that
			We can recognize in~\eqref{eq:proof_bound_omega_1c} the SINR at RX~$1$ for the centralized ZF scheme such that it holds:			
				\eqm{
						  {R_{1|\Omega}^{\HAP} (\alpha^{(1)}  ,\alpha^{(2)})} 
								&	 =  \ExpHCo{ \log_2 \bigg(   1 + \frac{\frac{P}{2}|{\hv_1^{\He}\tv^{\HAP}_1}|^2}{1 + \frac{P}{2}|{\hv_1^{\He}\tv^{\HAP}_2}|^2}  \bigg) }   \\ %\hspace{13ex}						
								& \geq -\ExpHCo{\log_2\LB\Gamma_1 \RB} + {R_{1|\Omega}^{\ZF}(\alpha^{(1)})}. \label{eq:proof_bound_omega_3}
				}		
			Since ${\DeltaR_1} \, =  {R_{1|\Omega}^{\ZF} \,( \,\alpha^{(1)} )} - {R_{1|\Omega}^{\HAP} ( \,\alpha^{(1)}  ,\alpha^{(2)})}$, it follows that
			\eqm{
					{\DeltaR_1} 
							%&= {R_{1|\Omega}^{\ZF}(\alpha^{(1)})} - {R_{1|\Omega}^{\HAP}(\alpha^{(1)},\alpha^{(2)})}  \\
							& \leq \ExpHCo{\log_2\LB\Gamma_1 \RB}. \label{eq:proof_bound_omega_4}
				}									
		%			
		%%%%%%%%%%%%%%%%%%%%%%%%%%%%%%%%%%%%%%%%%%%%%%%%%%%%%
		%%%%%%%%%%%%%%%%%%%%%%%%%%%%%%%%%%%%%%%%%%%%%%%%%%%%%
			%\subsection{Rate-offset conditioned on $\Omega^\setcomp$ (inconsistent precoding)}\label{subse:omegac_condition}
		%%%%%%%%%%%%%%%%%%%%%%%%%%%%%%%%%%%%%%%%%%%%%%%%%%%%%				
		%%%%%%%%%%%%%%%%%%%%%%%%%%%%%%%%%%%%%%%%%%%%%%%%%%%%%	
			%The rate ${R_{1|\Omega^\setcomp}^{\ZF}(\alpha^{(1)})}$  is trivially bounded by the interference-free single-user rate to obtain that 
				%\eqm{
						%{R_{1|\Omega^\setcomp}^{\ZF}(\alpha^{(1)})} 
								%& \leq \log_2\LB 1 + P\,\ExpHC{\norm{\hv_1}^2}\RB \\
								%& =    \log_2\LB 1 + 2P\RB.  \label{eq:proof_bound_omegac_2}
				%}
			Focusing  on the \emph{inconsistent precoding} cases,  ,  since $ {R_{1|\Omega^\setcomp}^{\HAP}(\alpha^{(1)},\alpha^{(2)})}\geq 0$ the rate gap can be bounded by the centralized rate as $\DeltaR_{1|\Omega^\setcomp} \leq {R_{1|\Omega^\setcomp}^{\ZF}(\alpha^{(1)})}$. % ${R_{1|\Omega^\setcomp}^{\HAP}(\alpha^{(1)},\alpha^{(2)})}\geq 0$   that  
				%
			%\noindent 
			%Focusing now on the \emph{inconsistent precoding} cases, i.e., $\ExpHCoc{\Delta r_1}$,  since $ {R_{1|\Omega^\setcomp}^{\HAP}(\alpha^{(1)},\alpha^{(2)})}\geq 0$  we can write that 
				%\eqm{
					%\ExpHCoc{\Delta r_1} 
							%%&= {R_{1|\Omega^\setcomp}^{\ZF}(\alpha^{(1)})} - {R_{1|\Omega^\setcomp}^{\HAP}(\alpha^{(1)},\alpha^{(2)})}  \\
							%& \leq {R_{1|\Omega^\setcomp}^{\ZF}(\alpha^{(1)})}. \label{eq:proof_bound_omegac_1}
				%}	
		%%%%%%%%%%%%%%%%%%%%%%%%%%%%%%%%%%%%%%%%%%%%%%%%%%%%%
		%%%%%%%%%%%%%%%%%%%%%%%%%%%%%%%%%%%%%%%%%%%%%%%%%%%%%
			%\subsection{Rate-offset Lower-bound}\label{subse:omegac_condition}
		%%%%%%%%%%%%%%%%%%%%%%%%%%%%%%%%%%%%%%%%%%%%%%%%%%%%%				
		%%%%%%%%%%%%%%%%%%%%%%%%%%%%%%%%%%%%%%%%%%%%%%%%%%%%%				
			%Introducing~\eqref{eq:proof_bound_omega_4} and~\eqref{eq:proof_bound_omegac_2} in~\eqref{eq:proof_3b} yields
			Putting these results together in~\eqref{eq:proof_3b}  yields
				\eqm{
						&\DeltaR_1 = {R_1^{\ZF}(\alpha^{(1)})}	-		 {R_1^{\HAP}(\alpha^{(1)},\alpha^{(2)})} \\
								& \ \leq \ExpHCo{2\log_2\LB\frac{\lambda^{(1)}_i}{\lambda^{\Qc}_i} \RB} +\Pr\LB\Omega^\setcomp\RB {R_{1|\Omega^\setcomp}^{\ZF}(\alpha^{(1)})},			\label{eq:proof_rate_bound_1}
				}
			where we have applied the fact that $\Pr\LB\Omega\RB\leq1$. Thus, since $\Gamma_1$ and $\Gamma_2$ are identically distributed, it holds that
				\eqm{
						&\DeltaR \leq 2\DeltaR_1,
				}			
			%Following the same approach, we can similarly show that $\DeltaR\geq-2\DeltaR_1$ (see extended version), 
			which concludes the proof. %. Consequently, Theorem~\ref{thm:bound_gap} is proven. 

	%%%%%%%%%%%%%%%%%%%%%%%%%%%%%%%%%%%%%%%%%%%%%%%%%%%%%
	%%%%%%%%%%%%%%%%%%%%%%%%%%%%%%%%%%%%%%%%%%%%%%%%%%%%%
	%%%%%%%%%%%%%%%%%%%%%%%%%%%%%%%%%%%%%%%%%%%%%%%%%%%%%
		\section{Numerical Results}\label{se:example}
	%%%%%%%%%%%%%%%%%%%%%%%%%%%%%%%%%%%%%%%%%%%%%%%%%%%%%
	%%%%%%%%%%%%%%%%%%%%%%%%%%%%%%%%%%%%%%%%%%%%%%%%%%%%%
	%%%%%%%%%%%%%%%%%%%%%%%%%%%%%%%%%%%%%%%%%%%%%%%%%%%%%			
		We illustrate in the following the performance for the uniform quantizer $\Qc_u$ introduced in  Lemma~\ref{lem:uniform_quantizer}.
		%, defined as
			%\eqm{
					%\Qc_u(x) = {\Pb^{\frac{-\alpha^{(2)}}{2}}}\big\lfloor \Pb^{\frac{\alpha^{(2)}}{2}}x\big\rfloor. \label{eq:agree_proof_0ab}
			%}
		%The parameter $k$ provides flexibility to the quantizer, such that its value impacts the trade-off between the probability of agreement  and the rate of convergence towards the aimed centralized power control.  
		%In the following we analyze the performance of this precoding scheme. The proof that~$\Qc_u$ satisfies  properties~\eqref{eq:cond_convergence} and~\eqref{eq:cond_agreement} is left for the extended version due to space restrictions.
		For sake of exposition, we assume a simple power normalization that ensures the per-TX power constraint. Let us introduce the precoding vector of TX~$j$ before normalization as $\vv_{\TX j} = [\vv_{1,j}^{},\vv_{2,j}^{}]^\Trans$, such that  the final precoder of TX~$j$ is $\tv_{\TX j} =  [\mu_1\vv_{1,j}^{},\mu_2\vv_{2,j}^{}]^\Trans$. Then,  $\mu_i$ is chosen  as
				\eqm{
						\qquad\mu_i \triangleq \frac{1}{\max(\norm{\vv_{\TX 1}},\norm{\vv_{\TX 2}})} \qquad \forall i\in\{1,2\}.\label{eq:def_mu}
				}			
		%%%%%%%%%%%%%%%%%%%%%%%%%%%%%%%%%%%%%%%%%%%%%%%%%%%%%
		%%%%%%%%%%%%%%%%%%%%%%%%%%%%%%%%%%%%%%%%%%%%%%%%%%%%%
			%\subsection{Performance analysis}\label{subse:performance}
		%%%%%%%%%%%%%%%%%%%%%%%%%%%%%%%%%%%%%%%%%%%%%%%%%%%%%
		%%%%%%%%%%%%%%%%%%%%%%%%%%%%%%%%%%%%%%%%%%%%%%%%%%%%%
			\begin{figure}[t] \centering%rate_k2_d1_a2_06
					\input{rate_TIT2012_a206_single.tikz}
							\caption{Expected sum rate of the proposed scheme for the setting with CSIT scaling parameters $\alpha^{(1)}=1$, $\alpha^{(2)}=0.6$, using the uniform quantizer of Lemma~\ref{lem:uniform_quantizer}.%
							%The proposed scheme performance is shown for $k=2$.
							}	\label{fig:fig_perf1}%
			\end{figure}
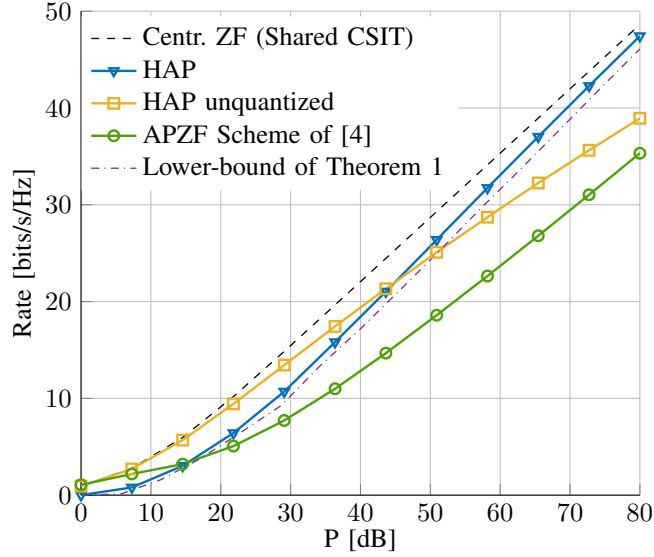
					
			In Fig.~\ref{fig:fig_perf1}, we simulate the expected sum rate of the proposed scheme using Monte-Carlo runs and averaging over 1000 random codebooks and 1000 channel realizations, for the CSIT configuration  $\alpha^{(1)}=1$ and $\alpha^{(2)}=0.6$. %The value of $k$ is chosen as $k=2$. 
						
			We can see that the proposed scheme leads to a vanishing loss with respect to the centralized case (where both TXs are provided with the best CSIT), and that the lower-bound of Theorem~\ref{thm:bound_gap} is considerably close to the actual rate. Furthermore, the scheme given in \cite{dekerret2012_TIT} using a scaled power normalization of $P/\log_2(P)$ --so as to guarantee a full DoF and an instantaneous power constraint  for the precoder ${\tv_{\TX j}}$-- can be seen to achieve also the optimal DoF although at the cost of a strong loss in rate offset.
			Finally, we can see how using an unquantized coefficient at TX2 leads to a loss in terms of DoF. %		observe the importance of the quantization of the coefficients $\lambda_i$, as omitting to use a quantizer leads to a loss in DoF.
			This occurs because, as aforementioned, the mismatches between the precoding coefficients of each TX break the orthogonality needed for the interference nulling. Thus, this scheme only achieves a DoF proportional to $\alpha^{(2)}$ instead of $\alpha^{(1)}$.  
			At intermediate SNR, this unquantized scheme  outperforms the proposed HAP precoding scheme. Yet, this is a consequence of our focus in this work towards analytical tractability and asymptotic analysis. Optimizing the precoder for finite SNR performance will allow to bridge the gap between the two schemes to obtain a scheme outperforming both of them. 
			%This follows from the fact that we have used a simple quantizer satisfying the asymptotic properties,  without optimizing it for the performance at finite SNR. The optimization of the quantization function is left for future works. 

	%\FloatBarrier	
 	  
	%%%%%%%%%%%%%%%%%%%%%%%%%%%%%%%%%%%%%%%%%%%%%%%%%%%%%
	%%%%%%%%%%%%%%%%%%%%%%%%%%%%%%%%%%%%%%%%%%%%%%%%%%%%%
	%%%%%%%%%%%%%%%%%%%%%%%%%%%%%%%%%%%%%%%%%%%%%%%%%%%%%
		\section{Conclusion}\label{se:conclusion}
	%%%%%%%%%%%%%%%%%%%%%%%%%%%%%%%%%%%%%%%%%%%%%%%%%%%%%
	%%%%%%%%%%%%%%%%%%%%%%%%%%%%%%%%%%%%%%%%%%%%%%%%%%%%%
	%%%%%%%%%%%%%%%%%%%%%%%%%%%%%%%%%%%%%%%%%%%%%%%%%%%%%	
		Considering a decentralized scenario where each TX has a  CSI with different SNR scaling accuracy, we have shown that there exists a linear precoding scheme that asymptotically recovers the rate of ZF precoding in the ideal centralized setting in which the best estimate is shared.
		%This result is surprising as the gap between the respective accuracies also increases boundlessly. 
		Going beyond the setting considered, we have shown how using a low rate quantization of some parameters (here the power normalization) in combination with a higher-accuracy distributed decision allows to reach coordination without loosing precision. 
		The extension of the results to more antennas and more users, as well as the optimization at finite SNR, are interesting and challenging research problems currently under investigation.
		%for arbitrary number of users, as well as for other scenarios besides the Distributed Network MIMO, are challenging topics of ongoing research. 
	%%%%%%%%%%%%%%%%%%%%%%%%%%%%%%%%%%%%%%%%%%%%%%%%%%%%%
	%%%%%%%%%%%%%%%%%%%%%%%%%%%%%%%%%%%%%%%%%%%%%%%%%%%%%
	%%%%%%%%%%%%%%%%%%%%%%%%%%%%%%%%%%%%%%%%%%%%%%%%%%%%%	

	%%%%%%%%%%%%%%%%%%%%%%%%%%%%%%%%%%%%%%%%%%%%%%%%%%%%%%%%%%%%%%%%%%%%%%%%%%%%%%%%%%%%%%%%%%%%%%%%%%%%%%%%%%
	%%%%%%%%%%%%%%%%%%%%%%%%%%%%%%%%%%%%%%%%%%%%%%%%%%%%%%%%%%%%%%%%%%%%%%%%%%%%%%%%%%%%%%%%%%%%%%%%%%%%%%%%%%
	%%%%%%%%%%%%%%%%%%%%%%%%%%%%%%%%%%%%%%%%%%%%%%%%%%%%%%%%%%%%%%%%%%%%%%%%%%%%
	%%%%%%%%%%%%%%%%%%%%%%%%%%%%%%%%%%%%%%%%%%%%%%%%%%%%%
	%%%%%%%%%%%%%%%%%%%%%%%%%%%%%%
	%%%%%%%%%%%%%%%
	%%%%%%%
		\begin{appendices}			
	%%%%%%%
	%%%%%%%%%%%%%%%
	%%%%%%%%%%%%%%%%%%%%%%%%%%%%%%
	%%%%%%%%%%%%%%%%%%%%%%%%%%%%%%%%%%%%%%%%%%%%%%%%%%%%%
	%%%%%%%%%%%%%%%%%%%%%%%%%%%%%%%%%%%%%%%%%%%%%%%%%%%%%%%%%%%%%%%%%%%%%%%%%%%%
	%%%%%%%%%%%%%%%%%%%%%%%%%%%%%%%%%%%%%%%%%%%%%%%%%%%%%%%%%%%%%%%%%%%%%%%%%%%%%%%%%%%%%%%%%%%%%%%%%%%%%%%%%%
	%%%%%%%%%%%%%%%%%%%%%%%%%%%%%%%%%%%%%%%%%%%%%%%%%%%%%%%%%%%%%%%%%%%%%%%%%%%%%%%%%%%%%%%%%%%%%%%%%%%%%%%%%%

		%%%%%%%%%%%%%%%%%%%%%%%%%%%%%%%%%%%%%%%%%%%%%%%%%%%%%%
		%%%%%%%%%%%%%%%%%%%%%%%%%%%%%%%%%%%%%%%%%%%%%%%%%%%%%%
			\section{Proof of Lemma~\ref{lem:uniform_quantizer}}\label{app:uniform_quantizer}
		%%%%%%%%%%%%%%%%%%%%%%%%%%%%%%%%%%%%%%%%%%%%%%%%%%%%%%
		%%%%%%%%%%%%%%%%%%%%%%%%%%%%%%%%%%%%%%%%%%%%%%%%%%%%%%	
			
		We prove~Lemma~\ref{lem:uniform_quantizer} by means of showing that it holds for a more general case. Specifically, we prove that~$\forall k>1$, the quantizer
			\eqm{
					\Qc_u(x) \triangleq {\Pb^{\frac{-\alpha^{(2)}}{k}}}\big\lfloor \Pb^{\frac{\alpha^{(2)}}{k}}x\big\rfloor, \label{eq:agree_proof_0abqsdqs}
			}		
		satisfies properties~\eqref{eq:cond_bounded},~\eqref{eq:cond_convergence} and~\eqref{eq:cond_agreement}. We first prove property~\eqref{eq:cond_convergence}. Afterward, we demonstrate~\eqref{eq:cond_agreement} and finally~\eqref{eq:cond_bounded}.
				%%%%%%%%%%%%%%%%%%%%%%%%%%%%%%%%%%%%%%%%%%%%%%%%%%%%%
					\subsection{ Proof of \eqref{eq:cond_convergence}: Convergence}\label{subse:convergence_IEEEproof}
				%%%%%%%%%%%%%%%%%%%%%%%%%%%%%%%%%%%%%%%%%%%%%%%%%%%%%
					In order to prove that $\Qc_u$ satisfies~\eqref{eq:cond_convergence}, i.e., that
						%\begin{align}
								%\lim_{P\rightarrow \infty}\mathcal{Q}_{u}(\lambda^{(j)}_i) = \lambda_i^{\star}, \quad \text{a.s.}\quad \forall i,j\in\Nb_2, 	\label{eq:IEEEproof_c2_1}
						%\end{align}
					%we  remind that the power constraint assumed in this section implies that $\lambda^{(1)}_i = \lambda_i^{\star}$. Then, we can rewrite~\eqref{eq:IEEEproof_c2_1}  as
						\begin{align}
								\lim_{P\rightarrow \infty}\mathcal{Q}_{u}(\lambda^{(j)}_i) - \lambda_i^{(1)} = 0  \quad \text{a.s.}\quad \forall i,j\in \{1,2\},	\label{eq:IEEEproof_c2_2}
						\end{align}			
					we demonstrate~\eqref{eq:IEEEproof_c2_2} for $j = 2$, as the case with $j=1$ is straightforwardly proved following the same derivation. 
					Let $\vv^{(j)}_{\bH}\in \Rb^{8\times 1}$ be the column vector obtained by stacking the real and imaginary parts of the elements of $\hat{\bH}^{(j)}$ one on top of another, such that 
						\eqm{
							\vv^{(j)}_{\bH} = 
									\begin{bmatrix}
												\Re\LB\hh^{(j)}_{11}\RB \\
												\Im\LB\hh^{(j)}_{11}\RB \\
												\dots \\
												\Im\LB\hh^{(j)}_{22}\RB 
									\end{bmatrix},
						}
					where $\Re(x)$ (resp. $\Im(x)$) denotes the real (resp. imaginary) part of $x\in\Cb$.						
					Using the Taylor's expansion of $\lambda^{(2)}_i$  centered in $\lambda^{(1)}_i$, and introducing the notation 
						\eqm{
								\vartheta \triangleq \!\LB\!\big(\vv^{(2)}_{\bH} - \vv^{(1)}_{\bH}\big)^{\!\Transpose} \ \nabla\Lambda_i\!\LB\vv^{(1)}_{\bH}\RB + o\big( \norm{\vv^{(2)}_{\bH} - \vv^{(1)}_{\bH}}\big)\!\RB, \label{eq:IEEEproof_c2_2b}
						}	
					where $\LB\nabla\Lambda_i\!\LB\cdot\RB\RB^{\Transpose}$ is the $i$ row of the Jacobian  Matrix $\bJ_{\Lambda}$,	we have that
						\eqm{
								\lambda^{(2)}_i - \lambda^{(1)}_i = \vartheta.  \label{eq:IEEEproof_c2_2a}
						}
					%where $\vartheta$ is given as in~\eqref{eq:IEEEproof_vanish_dis_6b} by
					From \eqref{eq:IEEEproof_c2_2a} and the definition of $\Qc_u$ in~\eqref{eq:agree_proof_0abqsdqs}, it follows that 		
						\begin{align}
								&\Qc_{u}(\lambda^{(2)}_i) - \lambda^{(1)}_i   =  \Pb^{\frac{-\alpha^{(2)}}{k}}   \left\lfloor  \Pb^{\frac{\alpha^{(2)}}{k}} (\lambda^{(1)}_i   + \vartheta) \right\rfloor   - \lambda^{(1)}_i  . \label{eq:IEEEproof_c2_5aa}
						\end{align}					
					Then, since  $c\LF\frac{1}{c}(x+y)\RF-x\leq y $, we obtain that
						\begin{align}
								 \Qc_{u}(\lambda^{(2)}_i) - \lambda^{(1)}_i  &\leq \vartheta. 		\label{eq:IEEEproof_c2_5b} 
						\end{align}				
					Similarly, since  $c\LF\frac{1}{c}(x+y)\RF-x \geq c\LF\frac{y}{c}\RF\geq y-c $, we can bound~\eqref{eq:IEEEproof_c2_5aa} from below as
						\begin{align}
						&\Qc_{u}(\lambda^{(2)}_i) - \lambda^{(1)}_i 
								\geq  \vartheta - \Pb^{\frac{-\alpha^{(2)}}{k}}. \label{eq:IEEEproof_c2_7b} 
						\end{align}					
					From \eqref{eq:IEEEproof_c2_5b} and \eqref{eq:IEEEproof_c2_7b}, it holds that it is sufficient to prove that  
						\eqm{
								\limpf \vartheta  = 0	\qquad	\text{a.s.}
						}
					to demonstrate that $\lim_{P\rightarrow \infty}\Qc_{u}(\lambda^{(2)}_i) = \lambda^{(1)}_i$ almost surely. To do so, we make use of the following lemma, whose proof is relegated to  Appendix~\ref{app:IEEEproof_almostsure}. 
						\begin{lemma}\label{lem:limit_difference}						
								Let $\hat{\bH}^{(j)}$, $\forall j\in \{1,2\}$, be a quantized version of the matrix ${\tilde{\bH}}$, such that each row vector $\thv^{\Herm}_i$, $\forall i\in \{1,2\}$, is quantized with $B^{(j)}=\alpha^{(j)}\log_2(P)$ bits.  Then,  it holds that 
									\eqm{
											\limpf \norm{\vv^{(2)}_{\bH}-\vv^{(1)}_{\bH}} = 0	\qquad	\text{a.s.}
									}				
						\end{lemma}					
					\noindent Since $\norm{\nabla\Lambda_i}\leq \norm{\bJ_{\Lambda}}\leq M_\bJ$, it holds that
						\eqm{
								\abs{\vartheta} \leq%	\abs{\big(\vv^{(2)}_{\bH} - \vv^{(1)}_{\bH}\big)^{\!\Transpose} \nabla\Lambda_i\!\LB\vv^{(1)}_{\bH}\RB } \leq 
								\norm{\vv^{(2)}_{\bH}-\vv^{(1)}_{\bH}} M_\bJ + \abs{o\big( \norm{\vv^{(2)}_{\bH} - \vv^{(1)}_{\bH}}\big)} 
						}
					and thus we obtain  from Lemma~\ref{lem:limit_difference} that 
						\eqm{
								\limpf \vartheta 	& = 0\qquad	\text{a.s.}
						}	
					Consequently, $\Qc_u$ satisfies~\eqref{eq:cond_convergence}.

				%%%%%%%%%%%%%%%%%%%%%%%%%%%%%%%%%%%%%%%%%%%%%%%%%%%%%
					\subsection{Proof of~\eqref{eq:cond_agreement}: Probability of agreement}\label{subse:agreement_IEEEproof}
				%%%%%%%%%%%%%%%%%%%%%%%%%%%%%%%%%%%%%%%%%%%%%%%%%%%%%
					We want to prove that $\Qc_u$ satisfies that
						\eqm{
									\Pr\LB\Omega^\setcomp\RB = o\LB\frac{1}{\log_2(P)} \RB,  \label{eq:agree_IEEEproof_0}				
						}
					where $\Pr\LB\Omega^\setcomp\RB = 1-\Pr\LB \forall i\in \{1,2\}\  \Qc\big(\lambda^{(1)}_i\big)=\Qc\big(\lambda^{(2)}_i\big) \in\Rb^+\RB$. Note that, for any two events $A$, $B$, it holds that 
						\eqm{
								1-\Pr(A\land B) \leq 1-\Pr(A) + 1-\Pr(B).
						}
					Since the probability of agreement for $\lambda_1$ is the same as for $\lambda_2$, we can write
						\eqm{
							\Pr\LB\Omega^\setcomp\RB \leq 2\LB 1 - \Pr\LB\mathcal{Q}_{u}(\lambda^{(1)}_1)= \Qc_{u}(\lambda^{(2)}_1)\in\Rb^+ \RB\RB.  \label{eq:agree_IEEEproof_0b}				
						}
					Moreover, it holds that
						\eqm{
								1 - \Pr\LB\mathcal{Q}_{u}(\lambda^{(1)}_1)= \Qc_{u}(\lambda^{(2)}_1)\in\Rb^+ \RB 
									 \leq \Pr\LB\mathcal{Q}_{u}(\lambda^{(1)}_1)\neq \Qc_{u}(\lambda^{(2)}_1)\RB + \Pr\LB \Qc_{u}(\lambda^{(1)}_1)=0  \RB. \label{eq:proof_bound_lemma_1}
						}
					Focusing on the last term of~\eqref{eq:proof_bound_lemma_1} it follows that
						\eqm{
								\Pr\LB \Qc_{u}(\lambda^{(1)}_1)=0  \RB
									  & =  \Pr\LB \lambda^{(1)}_1 \leq \Pb^{\frac{-\alpha^{(2)}}{k}}   \label{eq:proof_vanish_zero_case1}\RB\\
										& \leq f^{\max}_{\Lambda_1} \Pb^{\frac{-\alpha^{(2)}}{k}} \label{eq:proof_vanish_zero_case2}\\
										& = o\LB\frac{1}{\log_2(P)}\RB, \label{eq:IEEEproof_vanish_dis_3_new}
						}
					where~\eqref{eq:proof_vanish_zero_case1} follows from the step-size of $\Qc_u$, and~\eqref{eq:proof_vanish_zero_case2} follows from the bounded density assumption of~\eqref{eq:bound_pdf}.
					%We denote $N=\Pb^{\frac{\alpha^{(2)}}{3}}$ the number of quantization levels of $\Qc_u$. 
					
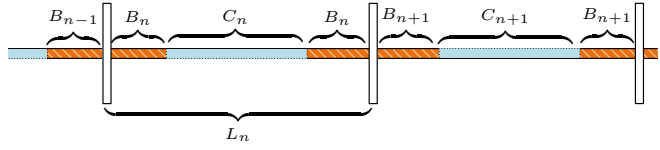
\begin{figure}[t]
							\centering
								\begin{overpic}[width=0.54\columnwidth]{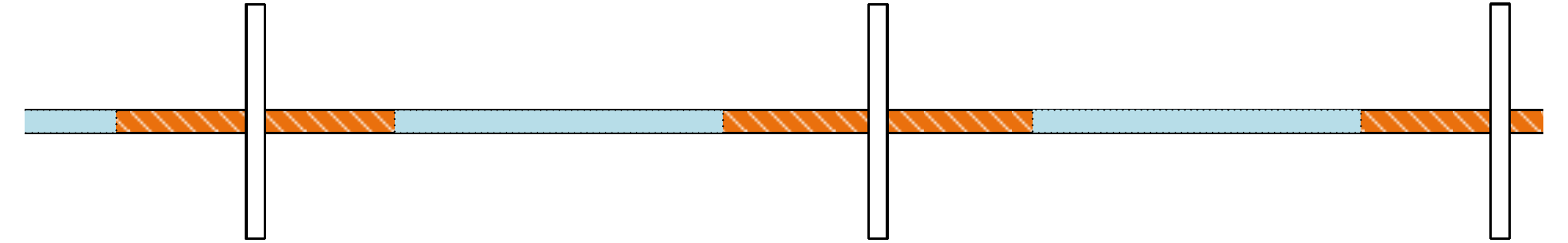}					
										{\put(16, -2.25){\parbox{\linewidth}{\small $\underbrace{\phantom{This is a really blank space}}_{L_n}$ }}}
										{\put(7.40, 9.5){\parbox{\linewidth}{\small $\overbrace{\phantom{ccici\,}}^{B_{n-1}}$ }}}
										{\put(17.0, 9.5){\parbox{\linewidth}{\small $\overbrace{\phantom{ccici\,}}^{B_n}$ }}}
										{\put(46.50, 9.5){\parbox{\linewidth}{\small $\overbrace{\phantom{ccici\;}}^{B_n}$ }}}
										{\put(25.5, 9.2){\parbox{\linewidth}{\small $\overbrace{\phantom{ o make space\,}}^{C_n}$ }}}
										{\put(65.75, 9.2){\parbox{\linewidth}{\small $\overbrace{\phantom{ o make space\,}}^{C_{n+1}}$ }}}
										{\put(57.0, 9.75){\parbox{\linewidth}{\small $\overbrace{\phantom{ccici\,}}^{B_{n+1}}$ }}}							
										{\put(87.0, 9.75){\parbox{\linewidth}{\small $\overbrace{\phantom{ccici\,}}^{B_{n+1}}$ }}}							
								\end{overpic}~\\ \vspace{1ex}  
								\caption{Illustration of a reconstruction level $L_n$ of the quantizer and the two sub-areas in which we divide it: The central  area $C_n$ and the edge area  $B_n$.} \label{fig:quantizer}
						\end{figure}

					Let us focus on the probability of disagreement $\Pr\big(\mathcal{Q}_{u}(\lambda^{(1)}_1)\neq \Qc_{u}(\lambda^{(2)}_1)\big)$. 				
					Let $\ell_n$ be the $n$-th reconstruction level of $\Qc_u$, $n\in\Nb_N \triangleq \{1,\dots,N\}$ with $N=\left\lceil \Pb^{\frac{\alpha^{(2)}}{k}}\right\rceil$. We assume that $\Pb^{\frac{\alpha^{(2)}}{k}}\in\Nb$ in order to ease the notation, although the result holds for any $\Pb^{\frac{\alpha^{(2)}}{k}}\in\Rb$. Let us define $L_n$ as the input interval that outputs $\ell_n$, i.e., 
					%$L_n = \{x\mid \Qc_u(x) = \ell_n\}$. $L_n$ 
						\eqm{
								L_n = \{x\mid \Qc_u(x) = \ell_n\}.
						}
					$L_n$ has a range $[\ell_n, \ell_{n+1})$ such that	$ \ell_{n+1} - \ell_n = \Pb^{\frac{-\alpha^{(2)}}{k}}$ ($\ell_{N+1} = 1$). We split $L_n$ in two areas, $B_n$ and $C_n$, depicted in~Fig.~\ref{fig:quantizer}. %, as depicted in Fig.~\ref{fig:quantizer}
					The area $B_n$ is the \emph{border} area, i.e.,
						\eqm{
							B_n \!=\!  \left\{\! x\in L_n\! \mid\! x - \ell_n^{\min} < \Pb^{\frac{-\alpha^{(2)}}{k-1}} \ \lor \   \ell_n^{\max} - x  < \Pb^{\frac{-\alpha^{(2)}}{k-1}}\!\!\right\}, 
						}
					whereas the area $C_n$ is the \emph{central} area, i.e.,	
						\eqm{
							C_n =  \left\{ x\in L_n \backslash B_n \right\}. 
						}		
					Intuitively, the probability of disagreement is very high if $\lambda^{(1)}_i$ lies in the border area $B_n$, whereas this probability vanishes in the central area $C_n$.
					Mathematically, we have that
						\eqm{
								&\Pr\LB\Qc_{u}(\lambda^{(1)}_1)\neq \Qc_{u}(\lambda^{(2)}_1) \RB \leq \Pr\LB\lambda^{(1)}_1\in \bigcup_{n\in\Nb_N} B_n \RB + \Pr\LB\Qc_{u}(\lambda^{(1)}_1)\neq \Qc_{u}(\lambda^{(2)}_1) \mid \lambda^{(1)}_1\in \bigcup_{n\in\Nb_N} C_n \RB\!.	\label{eq:IEEEproof_vanish_dis_1}
						}
					%%%%%%%%%%
					%%%%%%%%%%											
					From the bounded density assumption of~\eqref{eq:bound_pdf}, the probability that a computed value $\lambda^{(1)}_1$ is in $B_n$ is 
						\eqm{
								\Pr\LB\lambda^{(1)}_1\in B_n\RB &\leq f^{\max}_{\Lambda_1} |B_n| \\
										& = f^{\max}_{\Lambda_1} { 2\Pb^{\frac{-\alpha^{(2)}}{k-1}}}, 
						}
					where $|B_n|$ denotes the length of $B_n$. Since there are $N=\Pb^{\frac{\alpha^{(2)}}{k}}$ cells, the probability of being in the \emph{border} of any cell is
						\eqm{
								\Pr\LB\lambda^{(1)}_1\in \bigcup_{n\in\Nb_N} B_n\RB 
										& \leq \Pb^{\frac{\alpha^{(2)}}{k}}f^{\max}_{\Lambda_1} 2\Pb^{\frac{-\alpha^{(2)}}{k-1}} \\
										& = 2f^{\max}_{\Lambda_1}  \Pb^{\frac{-\alpha^{(2)}}{k(k-1)}} \label{eq:IEEEproof_vanish_dis_3_0}\\
										& = o\LB\frac{1}{\log_2(P)}\RB. \label{eq:IEEEproof_vanish_dis_3}
						}			
					Focusing on $C_n$, since  the minimum distance from any point of $C_n$ to the border of  $L_n$ is $\Pb^{\frac{-\alpha^{(2)}}{k-1}}$, it holds that
						\eqm{
								\Pr\LB\mathcal{Q}_{u}(\lambda^{(1)}_1)\neq \mathcal{Q}_{u}(\lambda^{(2)}_1) \ \Big|\  \lambda^{(1)}_1 \in \bigcup_{n\in\Nb_N} C_n\RB  
										& \leq \Pr\LB\big|{\lambda^{(1)}_1-\lambda^{(2)}_1}\big| \geq \Pb^{\frac{-\alpha^{(2)}}{k-1}}\ \Big|\  \lambda^{(1)}_1 \in \bigcup_{n\in\Nb_N} C_n\RB.  \label{eq:IEEEproof_quant_1a}
						}	
					Given that, for two events $A,C$, $\Pr(A\mid C) \leq \Pr(A)/\Pr(C)$, it follows that
						\eqm{
							\Pr\LB\big|{\lambda^{(1)}_1-\lambda^{(2)}_1}\big| \geq \Pb^{\frac{-\alpha^{(2)}}{k-1}} \ \Big|\  \lambda^{(1)}_1 \in \bigcup_{n\in\Nb_N} C_n\RB 
									&\leq \frac{1}{\Pr\LB\lambda^{(1)}_1 \in \bigcup_{n\in\Nb_N} C_n\RB}  \Pr\LB\abs{\lambda^{(1)}_1-\lambda^{(2)}_1} \geq \Pb^{\frac{-\alpha^{(2)}}{k-1}}\RB \\
								  &\leq \frac{1}{1-2f^{\max}_{\Lambda_1}  \Pb^{\frac{-\alpha^{(2)}}{k(k-1)}}}  \Pr\LB\abs{\lambda^{(1)}_1-\lambda^{(2)}_1} \geq \Pb^{\frac{-\alpha^{(2)}}{k-1}}\RB \label{eq:IEEEproof_quant_1aab}											 \\
									&\leq \frac{1}{1-2f^{\max}_{\Lambda_1}  \Pb^{\frac{-\alpha^{(2)}}{k(k-1)}}}  \frac{\Eb\left[\big|\lambda^{(1)}_1-\lambda^{(2)}_1\big|^2\right]}{\Pb^{\frac{-2\alpha^{(2)}}{k-1}}}, \label{eq:IEEEproof_quant_1b}											
						}	
					where~\eqref{eq:IEEEproof_quant_1aab} follows from~\eqref{eq:IEEEproof_vanish_dis_3_0} and~\eqref{eq:IEEEproof_quant_1b} from Chebyshev's Inequality.  
					In the following, we obtain the expectation  $\Eb\left[\big|\lambda^{(1)}_1-\lambda^{(2)}_1\big|^2\right]$. 
					From Taylor's Theorem it follows that%	From~\eqref{eq:IEEEproof_vanish_dis_6b},
						\eqm{
							\Eb\left[\big|\lambda^{(1)}_1-\lambda^{(2)}_1\big|^2\right] 
											&\leq \Eb\left[\left|\LB\vv^{(2)}_{\bH} - \vv^{(1)}_{\bH}\RB^{\Transpose} \nabla\Lambda_1\LB\vv^{(1)}_{\bH}\RB\right| ^2\right]  + \Eb\left[\left|o\LB \norm{\vv^{(2)}_{\bH} - \vv^{(1)}_{\bH}}\RB\right| ^2\right] \\
								& \leq  M^2_\bJ \Eb\left[\norm{\vv^{(2)}_{\bH} - \vv^{(1)}_{\bH}}^2\right]  	+ \Eb\left[o\LB \norm{\vv^{(2)}_{\bH} - \vv^{(1)}_{\bH}}\RB\right],  \label{eq:IEEEproof_vanish_dis_6ca} %%\\							
								%& \leq (M_\bJ + 1)E\left[\left|\norm{\vv^{(2)}_{\bH} - \vv^{(1)}_{\bH}}  \right| ^2\right] 		\label{eq:IEEEproof_vanish_dis_6cb}
						}			  
					where~\eqref{eq:IEEEproof_vanish_dis_6ca} comes from the fact that $ \norm{\nabla\Lambda_i}\leq \norm{\bJ_{\Lambda}}\leq M_\bJ$. % and~\eqref{eq:IEEEproof_vanish_dis_6cb} from 
						%\eqm{
								%E\left[o\LB \norm{\vv^{(2)}_{\bH} - \vv^{(1)}_{\bH}}\RB\right] \leq E\left[\left|\norm{\vv^{(2)}_{\bH} - \vv^{(1)}_{\bH}}  \right| ^2\right].
						%}
					We present in the following a useful lemma whose proof is relegated to Appendix~\ref{app:IEEEproof_expectation}.
						\begin{lemma}\label{lem:order_difference}
								Let $\hat{\bH}^{(j)}$, $\forall j\in \{1,2\}$, be a quantized version of the matrix $\tilde{\bH}$, such that each row vector $\tbh^{\Herm}_i$, $\forall i\in \{1,2\}$, is quantized with $B^{(j)}=\alpha^{(j)}\log_2(P)$ bits. Let $\kappa$ be a positive constant and $\alpha^{(1)} \geq \alpha^{(2)}$. Then, it holds that 
									\eqm{
											\Eb\left[\norm{\vv^{(2)}_{\bH} - \vv^{(1)}_{\bH}}^2\right] = \kappa P^{-\alpha^{(2)}}.
									}						
						\end{lemma}
					\noindent It follows from Lemma~\ref{lem:limit_difference} and Lemma~\ref{lem:order_difference} than 
						\eqm{
								\Eb\left[o\LB \norm{\vv^{(2)}_{\bH} - \vv^{(1)}_{\bH}}\RB\right] = o\big(P^{-\alpha^{(2)}}\big). \label{eq:bound_c2_oo}
						}
					Including Lemma~\ref{lem:order_difference} and~\eqref{eq:bound_c2_oo} in~\eqref{eq:IEEEproof_vanish_dis_6ca} yields
						\eqm{
						\Eb\left[\big|\lambda^{(1)}_1-\lambda^{(2)}_1\big|^2\right] 
							& \leq  \kappa M^2_\bJ  P^{-\alpha^{(2)}} + o\big( P^{-\alpha^{(2)}}\big).  \label{eq:IEEEproof_vanish_dis_6d} %%\\							
							%& \leq (M_\bJ + 1)E\left[\left|\norm{\vv^{(2)}_{\bH} - \vv^{(1)}_{\bH}}  \right| ^2\right] 		\label{eq:IEEEproof_vanish_dis_6cb}
						}
					Since $\Pb = \sqrt{P}$, substituting~\eqref{eq:IEEEproof_vanish_dis_6d} in~\eqref{eq:IEEEproof_quant_1b} we obtain that
						\eqm{
								{\Pr\LB\mathcal{Q}_{u}(\lambda^{(1)}_1)\neq \mathcal{Q}_{u}(\lambda^{(2)}_1)\mid \lambda^{(1)}_1 \in \bigcup_{n\in\Nb_N} C_n\RB}  
										& \leq \frac{1}{1-2f^{\max}_{\Lambda_1}  \Pb^{\frac{-\alpha^{(2)}}{k(k-1)}}}   \frac{\kappa M^2_\bJ P^{-\alpha^{(2)}} + o\LB P^{-\alpha^{(2)}}\RB}{P^{\frac{-\alpha^{(2)}}{k-1}}}, \label{eq:IEEEproof_vanish_dis_6e} \\
										&  = O\LB P^{-\alpha^{(2)}\frac{k-2}{k-1}}\RB\\
										&  = o\LB\frac{1}{\log_2(P)}\RB\label{eq:proof_lem_bound_p2_6}
						}				
					for any $k>1$. From~\eqref{eq:IEEEproof_vanish_dis_3_new},~\eqref{eq:IEEEproof_vanish_dis_3}~and~\eqref{eq:proof_lem_bound_p2_6} it follows that
						\eqm{
							\Pr\LB\Omega^\setcomp\RB 
									&\leq 2\LB \Pr\LB\mathcal{Q}_{u}(\lambda^{(1)}_1)\neq \Qc_{u}(\lambda^{(2)}_1)\RB + \Pr\LB \Qc_{u}(\lambda^{(1)}_1)=0  \RB\RB  \\
									& = o\LB\frac{1}{\log_2(P)}\RB,
						}					
					what concludes the proof for property~\eqref{eq:cond_agreement}.	
					
					%\newpage
					
				%%%%%%%%%%%%%%%%%%%%%%%%%%%%%%%%%%%%%%%%%%%%%%%%%%%%%
					\subsection{Proof of~\eqref{eq:cond_bounded}: Bounded Expectation}\label{subse:bounded_IEEEproof}
				%%%%%%%%%%%%%%%%%%%%%%%%%%%%%%%%%%%%%%%%%%%%%%%%%%%%%			
					We show in the following that $\exists M <\infty$ such that $\forall P$ it holds that 
						\eqm{
								\abs{\Eb_{\mid\Qc_u(\lambda^{(j)}_i)>0}\big[{\log_2\big({\Qc_u(\lambda^{(j)}_i)}\big)}\big]} \leq M. \label{eq:cond_bossunded} 
						}	
					Let us denote $x \triangleq \lambda^{(j)}_i$, $i,j\in\{1,2\}$ and the quantization step size as $q\triangleq \Pb^{-\alpha^{(2)}/k}$. 
					First, we easily upper bound it as $0\leq \lambda^{(j)}_i\leq 1$ implies that 
						\eqm{
								\Eb_{\mid{\Qc_u(x)}>0}\big[{\log_2\big({\Qc_u(x)}\big)}\big]\leq 0.
						}
					In order to lower bound it, note that
						\eqm{
								\Eb_{\mid\Qc_u(x)>0}\big[{\log_2\big({\Qc_u(x)}\big)}\big] \triangleq \sum_{i=1}^M \log_2(iq) p_{\mid \Qc_u(x)>0}(iq), \label{eq:proof_bound_mena_0}
						}
					where $M\triangleq \left\lceil \frac{1}{q}\right\rceil - 1$ because the quantization level $\Qc_u(x)=0$ ($i=0$) is excluded from $\Qc_u(x)>0$. 
					%For sake of readability, we  assume that $\frac{1}{q}\in\Nb$ such that $M = \frac{1}{q}-1$. 
					Besides this, the term $p_{\mid \Qc_u(x)>0}(iq)$ stands for 
						\eqm{
								p_{\mid \Qc_u(x)>0}(iq) \triangleq \Pr\LB\Qc_u(x) = iq \mid \Qc_u(x)>0\RB,
						}
					where  $ \Pr\LB\Qc_u(x) = iq \RB = \Pr\LB iq \leq x \leq (i+1)q\RB$.  
					The expectation in~\eqref{eq:proof_bound_mena_0} is bounded for a given finite $P$ because $q=\Pb^{-\alpha^{(2)}/k}>0$. In the following we prove that it is bounded also when $P\rightarrow\infty$.  We can write that
						\eqm{
								\Pr\LB\Qc_u(x) = iq \mid \Qc_u(x)>0\RB 
										&= \frac{\Pr\LB\Qc_u(x) = iq \land \Qc_u(x)>0\RB}{\Pr\LB \Qc_u(x)>0\RB} \\
										&= \frac{\Pr\LB\Qc_u(x) = iq \RB}{\Pr\LB \Qc_u(x)>0\RB} \\
										&\leq \frac{f^{\max}_{\Lambda}}{\max(0,1-f^{\max}_{\Lambda}\Pb^{\frac{-\alpha^{(2)}}{k}})}q, \label{eq:proof_bound_measssn_3}
						}	
					where~\eqref{eq:proof_bound_measssn_3} comes from~\eqref{eq:proof_vanish_zero_case2} as $1 - \Pr\LB \Qc_u(x)>0\RB \leq f^{\max}_{\Lambda} \Pb^{\frac{-\alpha^{(2)}}{k}}$ and from the fact that $\Pr\LB\Qc_u(x) = iq \RB\leq f^{\max}_{\Lambda}q$. 
					Note that $\exists P_{\min}$ such that $\forall P>P_{\min}$, $1-f^{\max}_{\Lambda}\Pb^{\frac{-\alpha^{(2)}}{k}} > 0$. As we focus on the limit as $P\rightarrow\infty$, we assume hereinafter that $1-f^{\max}_{\Lambda}\Pb^{\frac{-\alpha^{(2)}}{k}} > 0$.
					 We introduce the notation 
						\eqm{
								p'_{\max}\triangleq  \frac{f^{\max}_{\Lambda}}{1-f^{\max}_{\Lambda}\Pb^{\frac{-\alpha^{(2)}}{k}}}.
						}
					Hence, since $M\leq \frac{1}{q}$ (and thus $q\leq \frac{1}{M}$) and $\forall i\leq \frac{1}{q}$ it holds that $\log_2(iq)\leq 0$, it follows that
						\eqm{
							\Eb_{\mid\Qc_u(x)>0}\big[{\log_2\big({\Qc_u(x)}\big)}\big] 
									& \geq \sum_{i=1}^{M} \log_2\LB{iq}\RB {p'_{\max}q}  \label{{eq:proof_bound_mean_3}} \\
									& \geq \sum_{i=1}^{M} \log_2\LB\frac{i}{M}\RB \frac{p'_{\max}}{M}  \label{{eq:proof_bound_mean_3b}} \\
									& = \frac{p'_{\max}}{M}\LB \sum_{i=1}^{M}\log_2(i) - \sum_{i=1}^{M} \log_2(M)  \RB \\  %\label{{eq:proof_bound_mean_3}}
									& = {p'_{\max}}\LB  \frac{\log_2(M!)}{M} - \log_2(M)  \RB.  %\label{{eq:proof_bound_mean_3}}
						}					
					We have that 
						\eqm{
							\lim_{M\rightarrow\infty} \LB  \frac{\log_2(M!)}{M} - \log_2(M)  \RB = \frac{-1}{\ln(2)},  %\label{{eq:proof_bound_mean_3}}
						}					
					what together with the fact that $\limpf p'_{\max} = f^{\max}_{\Lambda}$ implies that
						\eqm{
							\limpf \Eb_{\mid\Qc_u(x)>0}\big[{\log_2\big({\Qc_u(x)}\big)}\big] 
									& \geq \frac{-f^{\max}_{\Lambda}}{\ln(2)}.  %\label{{eq:proof_bound_mean_3}}
						}						
					what concludes the proof. 
		%%%%%%%%%%%%%%%%%%%%%%%%%%%%%%%%%%%%%%%%%%%%%%%%%%%%%%
		%%%%%%%%%%%%%%%%%%%%%%%%%%%%%%%%%%%%%%%%%%%%%%%%%%%%%%
			\section{Proof of Lemma~\ref{lem:limit_difference}}\label{app:IEEEproof_almostsure}
		%%%%%%%%%%%%%%%%%%%%%%%%%%%%%%%%%%%%%%%%%%%%%%%%%%%%%%
		%%%%%%%%%%%%%%%%%%%%%%%%%%%%%%%%%%%%%%%%%%%%%%%%%%%%%%	
			Let us denote the first element of the vector $\vv^{(j)}_{\bH}\in\Rb^{{8\times 1}}$  as $\hh^{(j)}_{\Re}$, i.e., $\hh^{(j)}_{\Re} = \Re\big(\hh^{(j)}_{11}\big)$. Similarly, $\tilde{\h}_{\Re}$ denotes the real part of the normalized channel coefficient, $\tilde{\h}_{\Re} = \Re\big({\tilde{\h}}_{11}\big)$. Therefore, since the elements of $\vv^{(j)}_{\bH}$ are i.i.d.,
				\eqm{
							\norm{\vv^{(2)}_{\bH}-\vv^{(1)}_{\bH}} \overset{a.s.}{\longrightarrow} 0  \quad \Longleftrightarrow \quad  \abs{\hh^{(2)}_{\Re} - \hh^{(1)}_{\Re}}^2 \overset{a.s.}{\longrightarrow} 0.
				}
			Furthermore, from the feedback model it holds that
				\eqm{
							\abs{\hh^{(2)}_{\Re}\! - \hh^{(1)}_{\Re}}^2 \overset{a.s.}{\longrightarrow} 0  \  \Longleftrightarrow \   \hh^{(j)}_{\Re}\! - \tilde{\h}_{\Re} \overset{a.s.}{\longrightarrow} 0 \  \forall j\in \{1,2\}.
				}			
			Let $A_n=\left\{\abs{X_n - X}>\varepsilon\right\}$. Then,
				\eqm{
						X_n \overset{a.s.}{\longrightarrow} X  \quad \Longleftrightarrow  \quad \Pr\LB A_n\  i.o. \RB = 0 \quad \forall\varepsilon>0,
				}
			where 
				\eqm{
						A_n\  i.o. &\triangleq \left\{w:w\in A_n \text{ for infinitely many } n\right\}\\
											& = \limsup_n \ A_n. \label{eq:IEEEproof_lemma_3_1}
				}			
			Let $X_n = \hh^{(j)}_{\Re} - \tilde{\h}_{\Re}$ and $X= 0$. We obtain in the following $\Pr\LB A_n\RB =$ $\Pr\big( |\hh^{(j)}_{\Re} - \tilde{\h}_{\Re}| > \varepsilon\big)$.  
			%Following the same argument as in~\eqref{eq:IEEEproof_lemma_prob_4_a}-\eqref{eq:IEEEproof_lemma_prob_6}, it holds that
			The absolute value of the difference can be bounded as
				\eqm{
					\abs{\hh^{(j)}_{\Re} - \tilde{\h}_{\Re}} 
						& =  \abs{(1\!-\!\zop^{(j)}_1)\hh^{(j)}_{\Re} \! - z^{(j)}_1\delta_{\Re}^{(j)}} \label{eq:IEEEproof_lemma_prob_4y}\\
						& \leq  {(1-\zop^{(j)}_1)}+ {z^{(j)}_1} \label{eq:IEEEproof_lemma_prob_5y}
				}		
			where~\eqref{eq:IEEEproof_lemma_prob_4y} comes from the estimate model in~\eqref{eq:SM_err2} and~\eqref{eq:IEEEproof_lemma_prob_5y} because $|\hh^{(j)}_{\Re}| \leq 1$ and $|\delta_{\Re}^{(j)}| \leq 1$. The absolute value is omitted in~\eqref{eq:IEEEproof_lemma_prob_5y} because $0 \leq z^{(j)}_1\leq 1$. 
			Let us remind that %The term $\zop^{(j)}_1$ is defined as
				\eqm{
						\zop^{(j)}_1 = \sqrt{1-(z^{(j)}_1)^2}.
				}
			Since $1-\sqrt{1-x^2}\leq x$ for $0\leq x\leq 1$, it holds that
				\eqm{
						|{\hh^{(j)}_{\Re} - \tilde{\h}_{\Re}}|
								& \leq 2{z^{(j)}_1}. \label{eq:IEEEproof_lemma_prob_5z}
				}
			Hence,
				\eqm{
						\Pr\LB \abs{\hh^{(2)}_{\Re} - \tilde{\h}_{\Re}} > \varepsilon\RB 
								& \leq \Pr\LB  2{z^{(j)}_1} > \varepsilon\RB \\
								& = \Pr\LB  {Z^{(j)}_1} > \frac{\varepsilon^2}{4}\RB, 
				}
			since $Z^{(j)}_1 = (z^{(j)}_1)^2$. The quantization error $Z^{(j)}_1$ is distributed as the minimum of $n=2^{B^{(j)}}$ (and $2^{B^{(j)}}=P^{\alpha^{(j)}})$ standard uniform random variables \cite{Au-Yeung2007,Jindal2006}. Hence, upon denoting  $\varepsilon^\prime = \frac{\varepsilon^2}{4}$, it holds that 
				\eqm{
						\Pr\LB  {Z^{(j)}_1} > \varepsilon^\prime \RB = \LB 1-\varepsilon^\prime \RB^n. \label{eq:IEEEproof_lemma_prob_7}
				}			
			By definition --see~\eqref{eq:IEEEproof_lemma_3_1}--, $\Pr\LB A_n\  i.o. \RB$ satisfies
				\eqm{
					 \Pr\LB A_n\  i.o. \RB \leq \lim_{n\rightarrow\infty} \sum_{m=n}^{\infty} \Pr\LB A_n\RB.\label{eq:IEEEproof_lemma_prob_8}
				}			
			Introducing~\eqref{eq:IEEEproof_lemma_prob_7} in~\eqref{eq:IEEEproof_lemma_prob_8} yields
				\eqm{
						\Pr\LB A_n\  i.o. \RB &\leq \lim_{n\rightarrow\infty} \sum_{m=n}^{\infty} \LB 1-\varepsilon^\prime \RB^n \\
						%& = \lim_{n\rightarrow\infty} \sum_{m=0}^{\infty} \LB 1-\varepsilon^\prime \RB^n  -  \sum_{m=0}^{n-1} \LB 1-\varepsilon^\prime \RB^n \\
						%& = \lim_{n\rightarrow\infty}\frac{1}{1-\LB 1-\varepsilon^\prime\RB}  -   \frac{1- \LB 1-\varepsilon^\prime\RB^{n-1}}{1-\LB 1-\varepsilon^\prime\RB}  \label{eq:IEEEproof_lemma_prob_9} \\
						%& = \lim_{n\rightarrow\infty}\frac{1}{\varepsilon^\prime}  -   \frac{1- \LB 1-\varepsilon^\prime\RB^{n-1}}{\varepsilon^\prime}   \\
						& = \lim_{n\rightarrow\infty}\frac{\LB 1-\varepsilon^\prime\RB^{n-1}}{\varepsilon^\prime} \label{eq:IEEEproof_lemma_prob_9}  \\
						& = 0.
				}
			where~\eqref{eq:IEEEproof_lemma_prob_9} comes from the application of the geometric series' formula. This implies that $\Pr\LB A_n\  i.o. \RB = 0$ $\forall \varepsilon>0$, and thus Lemma~\ref{lem:limit_difference} is proven.

		%%%%%%%%%%%%%%%%%%%%%%%%%%%%%%%%%%%%%%%%%%%%%%%%%%%%%%
		%%%%%%%%%%%%%%%%%%%%%%%%%%%%%%%%%%%%%%%%%%%%%%%%%%%%%%
			\section{Proof of Lemma~\ref{lem:order_difference}}\label{app:IEEEproof_expectation}
		%%%%%%%%%%%%%%%%%%%%%%%%%%%%%%%%%%%%%%%%%%%%%%%%%%%%%%
		%%%%%%%%%%%%%%%%%%%%%%%%%%%%%%%%%%%%%%%%%%%%%%%%%%%%%%				
			In this section we prove Lemma~\ref{lem:order_difference}, i.e., that 
				\eqm{
						E\left[\norm{\vv^{(2)}_{\bH} - \vv^{(1)}_{\bH}}^2\right] \leq \kappa P^{-\alpha^{(2)}}. \label{eq:app_lem2_1}
				}	
			As defined in the previous appendix, let $\hh^{(j)}_{\Re} = \Re\big(\hh^{(j)}_{11}\big)$ and $\tilde{\h}_{\Re}=\Re\big(\tilde{\h}_{11}\big)$. We start by noting that, since the elements of $\vv^{(j)}_{\bH}\in\Rb^{{8\times 1}}$ are i.i.d., it holds that		
				\eqm{
						E\left[\norm{\vv^{(2)}_{\bH} - \vv^{(1)}_{\bH}}^2\right] 
							%& = \sum_{n=1}^{8}E\left[\abs{\vv^{(2)}_{\bH,n}-\vv^{(1)}_{\bH,n}}^2\right]   . \label{eq:IEEEproof_lemma_prob_3}						
							& = 8 E\left[|{\hh^{(2)}_{\Re} - \hh^{(1)}_{\Re}}|^2\right]. \label{eq:IEEEproof_lemma_prob_3}						
				}
			The absolute value of the difference can be bounded as
				\eqm{
					 \abs{\hh^{(2)}_{\Re} - \hh^{(1)}_{\Re}} 
						& \leq  \abs{\hh^{(2)}_{\Re} - \tilde{\h}_{\Re}} + \abs{\tilde{\h}_{\Re}-\hh^{(1)}_{\Re}} \label{eq:IEEEproof_lemma_prob_4_a}\\
						& \leq 2z^{(2)}_1 + 2{z^{(1)}_1}, \label{eq:IEEEproof_lemma_prob_6}
				}
			what follows from~\eqref{eq:IEEEproof_lemma_prob_5z}. 
			Since $z^{(2)}_1$ is drawn from the same distribution as $z^{(1)}_1$ but with higher variance, it holds that 
				\eqm{
					E\left[\big(2{z^{(2)}_1} + 2z^{(1)}_1\big)^2\right] & \leq E\left[\big(4z^{(2)}_1\big)^2\right] 
				}
			and consequently
				\eqm{
						E\left[\abs{\hh^{(2)}_{\Re} - \hh^{(1)}_{\Re}}^2\right] 
							& \leq 16 E\left[\big(z^{(2)}_1\big)^2\right] \\
							& \leq 16 P^{-\alpha^{(2)}},\label{eq:IEEEproof_lemma_prob_8aa}
				}
			where~\eqref{eq:IEEEproof_lemma_prob_8aa} is obtained from \cite[Lemma~1]{Jindal2006}. %and comes from the fact that $\big(z^{(2)}_1\big)^2$ is the quantization error resulted from RVQ of the unit-norm channel vector.
			This concludes the proof of Lemma~\ref{lem:order_difference}.

	%%%%%%%%%%%%%%%%%%%%%%%%%%%%%%
	%%%%%%%%%%%%%%%
	%%%%%%%
		\end{appendices}			
	%%%%%%%
	%%%%%%%%%%%%%%%
	%%%%%%%%%%%%%%%%%%%%%%%%%%%%%%	

	%%%%%%%%%%
	%%%%%%%%%%%%%%%%%%%%
	%%%%%%%%%%%%%%%%%%%%%%%%%%%%%%%%%%%%%%%%%%%%
	%%%%%%%%%%%%%%%%%%%%%%%%%%%%%%%%%%%%%%%%%%%%%%%%%%%%%%%%%%%%%%%%%%%%%%
	%%%%%%%%%%%%%%%%%%%%%%%%%%%%%%%%%%%%%%%%%%%%%%%%%%%%%%%%%%%%%%%%%%%%%%%%%%%%%%%%%%%%
	%%%%%%%%%%%%%%%%%%%%%%%%%%%%%%%%%%%%%%%%%%%%%%%%%%%%%%%%%%%%%%%%%%%%%%%%%%%%%%%%%%%%%%%%%%%%%%%%
	%%%%%%%%%%%%%%%%%%%%%%%%%%%%%%%%%%%%%%%%%%%%%%%%%%%%%%%%%%%%%%%%%%%%%%%%%%%%%%%%%%%%%%%%%%%%%%%%
			\bibliographystyle{IEEEtran}												%%%%%%%%%%%%%%%%%%%%%%             %%%%%%%%%%%%%%%%%%%%%%
			\bibliography{./Literature}													%%%%%%%%%%%%%%%%%%%%%%             %%%%%%%%%%%%%%%%%%%%%%
	%%%%%%%%%%%%%%%%%%%%%%%%%%%%%%%%%%%%%%%%%%%%%%%%%%%%%%%%%%%%%%%%%%%%%%%%%%%%%%%%%%%%%%%%%%%%%%%%
	%%%%%%%%%%%%%%%%%%%%%%%%%%%%%%%%%%%%%%%%%%%%%%%%%%%%%%%%%%%%%%%%%%%%%%%%%%%%%%%%%%%%%%%%%%%%%%%%
	%%%%%%%%%%%%%%%%%%%%%%%%%%%%%%%%%%%%%%%%%%%%%%%%%%%%%%%%%%%%%%%%%%%%%%%%%%%%%%%%%%%%
	%%%%%%%%%%%%%%%%%%%%%%%%%%%%%%%%%%%%%%%%%%%%%%%%%%%%%%%%%%%%%%%%%%%%%%
	%%%%%%%%%%%%%%%%%%%%%%%%%%%%%%%%%%%%%%%%%%%%
	%%%%%%%%%%%%%%%%%%%%
	%%%%%%%%%%

\end{document}

%% file: Definitions.tex
\newcommand\itemEq[1][]{%
  \ifx\relax#1\relax  \item \else \item[#1] \fi
  \abovedisplayskip=0pt\abovedisplayshortskip=0pt~\vspace*{-\baselineskip}}
	
\newcommand{\itb}{\begin{itemize}}
\newcommand{\ite}{\end{itemize}}
\newcommand{\enb}{\begin{enumerate}}
\newcommand{\ene}{\end{enumerate}}
\newcommand{\eqm}[1]{\begin{align}#1\end{align}}

\newcommand{\limpf}{\lim_{P\rightarrow\infty}}
\newcommand{\setcomp}{{\mathsf{c}}}	
\newcommand{\Trans}{{\mathrm{T}}}

\newcommand{\LF}{\left\lfloor}
\newcommand{\RF}{\right\rfloor}
\newcommand{\LB}{\left(}
\newcommand{\RB}{\right)}
\newcommand{\LSB}{\left[}
\newcommand{\RSB}{\right]}

\newcommand{\abs}[1]{\left|#1\right|}

\newcommand{\h}{{\mathrm{h}}}
\newcommand{\hh}{\hat{{\mathrm{h}}}}

\newcommand{\bhh}{\hat{\hv}}

\newcommand{\tbh}{\tilde{\bh}}
\newcommand{\thv}{\tilde{\hv}}

\newcommand{\zop}{\breve{z}}

\newcommand{\expj}{^{(j)}}
\newcommand{\expo}{^{(1)}}
\newcommand{\expt}{^{(2)}}

\newcommand{\Pb}{{\bar{P}}}

\newcommand{\DeltaR}{\Delta\mathrm{R}}

\newcommand{\ExpB}[1]{\mathbb{E}\!\LSB#1\RSB}	
	
\newcommand{\ExpHC}[1]{\mathbb{E}\!\LSB#1\RSB}	
\newcommand{\ExpHCq}[1]{\mathbb{E}_{|\Qc(x)>0}\!\LSB#1\RSB\!}
\newcommand{\ExpHCo}[1]{\mathbb{E}_{|\Omega}\!\LSB#1\RSB\!}

\DeclareMathAlphabet{\mathbit}{OML}{cmr}{bx}{it}
\DeclareMathAlphabet{\mathsf}{OT1}{cmss}{m}{n}
\DeclareMathAlphabet{\mathTXf}{OT1}{cmss}{bx}{it}

\DeclareMathOperator{\Transpose}{T}

\DeclareMathOperator*{\argmin}{argmin}
\DeclareMathOperator*{\argmax}{argmax}

\DeclareMathOperator{\CN}{\mathcal{N}_{\mathbb{C}}}

\DeclareMathOperator{\DoF}{DoF}

\DeclareMathOperator{\TX}{TX}

\DeclareMathOperator{\ZF}{ZF}

\newcommand{\HAP}{\mathrm{HAP}}

\DeclareMathOperator{\SNR}{SNR}

\newcommand{\Nb}{{{\mathbb{N}}}}

\newcommand{\Rb}{{{\mathbb{R}}}}
\newcommand{\Cb}{{{\mathbb{C}}}}

\newcommand{\Eb}{{{\mathbb{E}}}}

\newcommand{\hv}{\mathbf{h}}

\newcommand{\tv}{\mathbf{t}}

\newcommand{\vv}{\mathbf{v}}

\newcommand{\xv}{\mathbf{x}}

\newcommand{\Cc}{{{\mathcal{C}}}}

\newcommand{\Lc}{{{\mathcal{L}}}}

\newcommand{\Qc}{{{\mathcal{Q}}}}

\newcommand{\bH}{\mathbf{H}}

\newcommand{\bJ}{\mathbf{J}}

\newcommand{\bT}{\mathbf{T}}

\newcommand{\bV}{\mathbf{V}}

\newcommand{\bh}{\bm{h}}

\newcommand{\bs}{\bm{s}}

\newcommand{\bv}{\bm{v}}
\newcommand{\bw}{\bm{w}}
\newcommand{\bx}{\bm{x}}
\newcommand{\by}{\bm{y}}

\newcommand{\Herm}{{{\mathrm{H}}}}

\newcommand{\norm}[1]{\lVert{#1}\rVert}

\newcommand{\He}{{{\mathrm{H}}}}

%% file: rate_TIT2012_a206_single.tikz
\definecolor{mycolor1}{rgb}{0.00000,0.44700,0.74100}%
\definecolor{mycolor2}{rgb}{0.85000,0.32500,0.09800}%
\definecolor{mycolor3}{rgb}{0.92900,0.69400,0.12500}%
\definecolor{mycolor4}{rgb}{0.49400,0.18400,0.55600}%
\definecolor{mycolor5}{rgb}{0.300000,0.64314,0.00000}%

\begin{tikzpicture}

\begin{axis}[%
			width=0.45\columnwidth,
			height=.39\columnwidth,
			at={(0\columnwidth,0\columnwidth)},
			scale only axis,
			xmin=0,
			xmax=80,
			xlabel={P [dB]},
			xmajorgrids,
			x label style={at={(axis description cs:0.5,0.03)},anchor=north},
			ymin=0,
			ymax=50,
			ylabel={Rate [bits/s/Hz]},
			ymajorgrids,
			y label style={at={(axis description cs:0.1,.5)},rotate=0,anchor=south},						
			axis background/.style={fill=white},
			title style={font=\bfseries},
			axis x line*=bottom,
			axis y line*=left,
			legend style={at={(0.001,0.63)},anchor=south west,legend cell align=left,align=left,draw=white}
]
\addplot [color=black,dashed,line width=0.5pt]
  table[row sep=crcr]{%
0	0.936158910063248\\
7.27272727272727	2.76341858657415\\
14.5454545454545	5.95734841345459\\
21.8181818181818	10.187195267692\\
29.0909090909091	14.8557792822771\\
36.3636363636364	19.6507432367861\\
43.6363636363636	24.4650354409531\\
50.9090909090909	29.2973861052496\\
58.1818181818182	34.1202723199756\\
65.4545454545455	38.9585574434502\\
72.7272727272727	43.7878927858119\\
80	48.6196795274233\\
};
\addlegendentry{Centr. ZF (Shared CSIT)\!\!\!\!};

\addplot [color=mycolor1,solid,line width=0.9pt,mark=triangle,mark options={solid,rotate=180}]
  table[row sep=crcr]{%
0	3.55650000419461e-07\\
7.27272727272727	0.795575811628086\\
14.5454545454545	3.03045682728602\\
21.8181818181818	6.4186013857615\\
29.0909090909091	10.7028984984568\\
36.3636363636364	15.8165690486539\\
43.6363636363636	21.0370836318893\\
50.9090909090909	26.3777713206408\\
58.1818181818182	31.7438970821344\\
65.4545454545455	37.0358892564175\\
72.7272727272727	42.276906295949\\
80	47.4291157780905\\
};
\addlegendentry{HAP};

\addplot [color=mycolor3,solid,line width=0.9pt,mark=square,mark options={solid}]
  table[row sep=crcr]{%
0	0.933372377067223\\
7.27272727272727	2.70576317680171\\
14.5454545454545	5.68694420719211\\
21.8181818181818	9.43048727726839\\
29.0909090909091	13.4360422626747\\
36.3636363636364	17.4340883260487\\
43.6363636363636	21.3235564815081\\
50.9090909090909	25.0913415652446\\
58.1818181818182	28.7174863351959\\
65.4545454545455	32.2508303575244\\
72.7272727272727	35.6267038950493\\
80	38.9369626163692\\
};
\addlegendentry{$\text{HAP unquantized}$};

\addplot [color=mycolor5,solid,line width=0.9pt,mark=o,mark options={solid}]
  table[row sep=crcr]{%
0	1.0734070922181\\
7.27272727272727	2.18657431414426\\
14.5454545454545	3.20552522070864\\
21.8181818181818	5.06600888180155\\
29.0909090909091	7.71625939627655\\
36.3636363636364	10.9897299594126\\
43.6363636363636	14.6719940339003\\
50.9090909090909	18.5883782300741\\
58.1818181818182	22.6385115533311\\
65.4545454545455	26.8003630716703\\
72.7272727272727	31.0359255954148\\
80	35.3243698739457\\
};
\addlegendentry{APZF Scheme of \cite{dekerret2012_TIT}};

\addplot [color=mycolor4,dashdotted, line width = 0.5pt]
  table[row sep=crcr]{%
0	0\\
5.71428571428571	0.122105682838121\\
11.4285714285714	1.59393179184614\\
17.1428571428571	3.79892028579227\\
22.8571428571429	6.47637940583235\\
28.5714285714286	9.22118205965463\\
34.2857142857143	13.4013048184225\\
40	17.1858395869366\\
45.7142857142857	21.2304335418374\\
51.4285714285714	25.330179325081\\
57.1428571428571	29.5711487846596\\
62.8571428571429	33.6355205566005\\
68.5714285714286	37.7989205264019\\
74.2857142857143	41.9797003617874\\
80	46.0930124451597\\
};
\addlegendentry{Lower-bound of Theorem~\ref{thm:bound_gap}\vspace{-5ex}};

\end{axis}
\end{tikzpicture}%